\documentclass[
preprint,
floatfix,
numerical 
]{jasatex}

\usepackage{moreverb}
\usepackage{citesort}
\usepackage{pst-solides3d,pgfplots}
\usepackage[caption]{subfig}
\usepackage{amsmath,amssymb}
\usepackage{setspace}
\usepackage{algorithm,algorithmic}
\graphicspath{{png//}}

\newtheorem{definition}{Definition}
\newcounter{defcnt}
\renewenvironment{definition}[1]{ {\textit {\\Definition \arabic{defcnt}: #1\\}}
}{%
\stepcounter{defcnt}
}
\newcommand{\norm}[2]{\left\Vert #2 \right\Vert_{#1}}  
\newcommand{\spn}[2]{span_#1 \left\{#2\right\}}

\begin{document}

\preprint{AIP/123-QED}

\title[Generalized Method of Moments]{Generalized Method of Moments: A novel framework for analyzing acoustic scattering from complex objects using a {\em locally} smooth surface parametrization and adaptive basis spaces}
\author{N. V. Nair}
\email{nairn@msu.edu}
\author{B. Shanker}%
 \altaffiliation[Also with]{Department of Physics and Astronomy, Michigan State University, East Lansing, MI 48824, USA.}
 \email{bshanker@egr.msu.edu}
 \author{L. Kempel}
 \email{kempel@egr.msu.edu}
\affiliation{%
Department of Electrical and Computer Engineering, Michigan State University, East Lansing, MI 48824, USA.}

\date{\today}

\begin{abstract}
The analysis of scattering from complex objects using surface integral equations is a challenging problem. Its resolution has wide ranging applications- from crack propagation to diagnostic medicine. The two ingredients of any integral equation methodology is the representation of the domain and the design of approximation spaces to represent physical quantities on the domain. The order of convergence depends on both the surface and geometry representation. For instance, most surface models are restricted to piecewise flat or second order tessellations. Similarly, the most commonly known basis spaces for acoustics are piecewise constant functions. What is desirable is a framework that permits adaptivity (of size and order) in both geometry and function representations. Unlike volumetric, differential equation solvers, such as the finite element method, developing an hp−adaptive framework for surface integral equations is very difficult. This papers proposes a resolution to this problem by developing a novel framework that relies on reconstruction of the surface using locally smooth parameterizations, and defining partition of unity functions and higher order basis spaces on overlapping domains. This permits easy refinement of both the geometry and function representation. This capabilities of the proposed framework are shown via a number of numerical examples.
\end{abstract}

\maketitle

\section{Introduction}\label{intro}

The analysis of scattering from complex objects has wide spread applicability, from crack propagation \cite{Koller1992,Portela1993} to non-destructive evaluation \cite{Bostrom2003,Maurice2007} to imaging and diagnostic medicine \cite{Caorsi2004,Kowar2010} to holography \cite{Chappell2009} to scattering from rough surfaces \cite{Agarwal1976,Colliander2007}, etc. Boundary integral formulations offer an efficient modality for the analysis of fields scattered by homogeneous objects as (i) they can be formulated only in terms of surface integral equations and (ii) radiation boundary conditions are explicitly included in the Green's function. Despite their advantages, their formulation is more difficult than that of their differential equation counterparts, and as a result this method has seen sporadic development in the past \cite{Schenck1967,Burton1971}, and a more concerted effort recently \cite{Bleszynski2008,Kawai2006,Makarov1998,Seydou2003,Wang2006,Willoughby2008,Yang2002,Yang2004}. The recent development of fast solvers that ameliorate the CPU and memory complexity of surface integral equation based solvers, i.e., reduce the scaling from ${\cal O}(N_s^2)$ to ${\cal O}(N_s \log_2 N_s)$ where $N_s$  is the number of spatial degrees of freedom, has made these techniques more appealing \cite{Coifman1993,Shanker2007}. However, when  compared to their differential equation based counterparts, the analysis here has been more or less restricted to simple basis functions (piecewise constant) and linear tessellations of the geometry. This is not to say that there is not a need for higher order function and geometric representation \cite{Schwab1998,Melenk1996,Babuska1997,Babuska2002}. Indeed, \cite{Babuska1997} makes an eloquent case for the development and use of such methods for scalar finite element problems. In this paper, our objective is to develop a flexible framework such that both the surface and function representations lend themselves to adaptivity in terms of patch size ($h-$), surface order ($g-$) and polynomial basis order ($p-$). In what follows, we shall review extant literature and motivate the need for such a method. 

Constructing an underlying mesh/tessellation is, perhaps, the most understated task in any computational analysis. Commercial meshing software exists that provides higher order tessellations, upto second order representation of surfaces. These meshes, while adequate for many applications, present challenges when one requires $h-$, $p-$ or $g-$adaptivity to ensure convergence of the solution or flexibility (mesh adaptation) to model crack propagation or deformation or the ability to handle non-conformal meshes (meshes that contain hanging nodes). Note, that the construction of a framework for $h-$, $p-$ and $g-$adaptivity for volumetric tessellations, such as those used in finite element methods is comparatively easy. It is apparent that a possible way to accomplish these goals is to define the topology of the scatterer using point clouds, and then use this to create local surface descriptions. Using point clouds to define a mesh is not new; it has been extensively studied with regard to efficiency in two dimensions and ${\cal O}(N\log N)$ algorithms exist \cite{Bowyer1981,Watson1981,Fournier1984,Toussaint1984,Seidel1991}. In three dimensions, creating a surface description using a set of point clouds is a highly overdetermined problem and therefore, considerably harder. A very successful algorithm to create tessellation from a set of points is the ball pivoting algorithm (BPA) \cite{Bernardini1999}. However, while the BPA has some deficiencies, the fundamental problem with this approach is that it creates a mesh. Once this mesh is created, it suffers from the same lack of adaptivity and flexibility alluded to earlier. Our approach to solving this problem is to develop a methodology that relies on defining piecewise overlapping domains. A local polynomial representation of these domains is obtained by ensuring that this representation matches the surface at a dense set of points specified by the point cloud. Given a local representation in terms of analytic functions, mechanisms are then developed to divide or merge patches ($h-$adaptivity), or change the local polynomial order of the surface ($g-$adaptivity), or both. Once a surface parametrization is obtained, the next step is the construction of function approximations on these surfaces. To this end, the two most desirable attributes of the surface parametrization are that: \begin{enumerate}
  \item It be local in nature. In particular, the parametrization should lend itself to the definition of \emph{local} approximation spaces
  \item It permits easy definition of surface derivatives and differential forms. All integral equation formulations require the construction of integrals on the surface and some formulations require multiple surface derivatives of functions. Thus, it is desirable that the surface parametrization allow for the construction of surface derivatives and Jacobians in closed-form. 
\end{enumerate} Our surface parametrization scheme draws inspiration from algorithms that exist in computer graphics \cite{Catmull1978,Grimm1995,Grimm2002} and computational physics  \cite{Bruno2007} literature.  However, to our knowledge, the algorithm presented here is the first that satisfies both the properties described above.

The other critical component of scattering analysis is the approximation space. As alluded to in \cite{Babuska1997}, it is eminently desirable to use higher order functions  for approximation, or better still, use functions that are based on the known local physics/heuristics. As is well known, these methods produce higher rates of convergence \cite{Schwab1998,Babuska1997}. Further, these features would reduce computational cost without the detrimentally affecting accuracy. However, it is challenging to mix different orders of basis functions, for instance, when one desires $p-$adaptivity or the use physically relevant basis sets in each region.  Incorporating such flexibility into classical solvers is very difficult as one has to take steps to ensure continuity of the physical quantity being represented. In the finite element community, the need/desire to enrich the approximation space, as well as ensure continuity gave rise to the generalized finite element method \cite{Melenk1996,Babuska1997,Chuan2007,Tuncer2010}, and its variations \cite{Duarte1996,Duarte2000,Griebel2002a}. The basis functions developed within this framework are continuous across domains and, as a result, do not need additional constraints to ensure continuity. The authors have recently developed methods that extend this idea to surface integral equations as applied to electromagnetics \cite{Nair2011,Nair2011a}, and have demonstrated convergence, well conditioned properties as well as application analysis of scattering from the range of targets. Unfortunately, this method still relies on an underlying tessellation.  While Nystr\"{o}m based schemes that use a point cloud and collocation, instead of a mesh and basis functions, are available \cite{Canino1998,Gedney2000}, these methods have been shown to be equivalent to higher order basis function schemes on a standard tessellation \cite{Peterson2002,Gedney2003}.  As a result, both of these methods suffer from the drawbacks alluded to in the earlier paragraph. 

The goal of this paper is to introduce a very general and flexible framework for surface integral equation based analysis of surface scattering. We will consider acoustic scattering from sound-hard objects as the target problem in this work. The method, called the Generalized Method of Moments (GMM), aims to introduce this range of functionality.  Specifically, in this paper we will
\begin{enumerate}
\item present the GMM computational framework that will 
\begin{enumerate}
\item introduce a framework to develop local analytical surface representations on overlapping domains starting from either a tesselated object or a point cloud
\item develop the mechanisms necessary for either merging or partitioning subdomains
\item develop the mechanisms for locally increasing/decreasing the order of representation of each subdomain
\item define basis functions with a partition of unity framework that are defined on these overlapping domains
\item describe accurate evaluation of integrals and the Galerkin solution process
\end{enumerate}
\item and present results that demonstrate 
\begin{enumerate}
\item  $h-$, $p-$, and $g-$convergence of surfaces using overlapping subdomians 
\item  convergence of function representation 
\item  convergence of scattering cross-section from canonical geometries and scattering cross-sections from topologically different objects
\item the ability of the method to mix different basis functions (polynomial or non-polynomial) in different regions or basis function adaptivity
\item $h$,$p-$, $hp-$, and $g-$adaptivity (in both surface and basis functions)
\end{enumerate} 
\end{enumerate}

Note, while it is not a direct focus of this work, the GMM framework introduced here can be easily accelerated using the fast multipole method \cite{Coifman1993,Shanker2007} to permit the analysis of very large objects.  The framework constructed here also permits easy integration with the fast multipole method  The rest of this paper proceeds as follows: In the next section, we will formally state the scattering problem. In Section \ref{sec:SSA}, we will describe the construction of the locally smooth surface parametrizations using either an underlying mesh or a point cloud. Section \ref{sec:GMMBas} details the construction of the basis functions on these surface parametrizations. The specifics of construction of the matrix elements will be elucidated in Section \ref{sec:mateval} and Section \ref{sec:res} will present several results that demonstrate the surface reconstruction, validate the basis function framework and showcase  the advantages of the proposed technique.  Finally, Section \ref{sec:concl} will provide some concluding remarks. 

\section{Problem Statement}
Let $D^-$ denote a rigid scatterer in a homogeneous medium bounded by $\Omega$ with a unique, outward pointing normal $\hat{\bf n}({\bf r}) \forall {\bf r} \in \Omega$. Consider a velocity field incident on this scatterer denoted by  ${\bf v}^i({\bf r})$.  This generates a scattered velocity field given by ${\bf v}^s({\bf r})$ and we define the total velocity as ${\bf v}^t({\bf r}) \doteq {\bf v}^i({\bf r}) + {\bf v}^s({\bf r})$. These fields can be represented by an equivalent potentials $\phi^\zeta({\bf r})$, $\zeta \in \{i,s,t\}$, where ${\bf v}^\zeta{\bf r}) \doteq \nabla \phi^\zeta({\bf r})$. Further, the corresponding pressure fields are given by $p^\zeta({\bf r}) \doteq -j\omega\rho_0\phi^\zeta({\bf r})$ where  $\rho_0$ is the density of the ambient medium. The total potential $\phi^t({\bf r}) = \phi^i({\bf r}) + \phi^s({\bf r})$ satisfies the Helmholtz equation and boundary condition given by 
\begin{equation}
  \begin{split}
    \nabla^2\phi^t({\bf r}) + k^2 \phi^t({\bf r}) &= 0 \qquad \forall \qquad{\bf r} \in {\mathcal R}^3/D^- \\
    \hat{\bf n}({\bf r}) \cdot \nabla\phi^t({\bf r}) &= 0 \qquad \forall\qquad {\bf r} \in \Omega.
\end{split}
  \label{eq:helm}
\end{equation}
The Kirchoff-Helmholtz integral theorem relates the scattered potential $\phi^s({\bf r})$ to the total potential as \begin{equation}
  \phi^s({\bf r}) = \int_\Omega d{\bf r} \phi^t({\bf r'}) \hat{\bf n}'({\bf r'}) \cdot \nabla' g({\bf r}, {\bf r}'), 
  \label{eq:kirchoff}
\end{equation}
where $g({\bf r},{\bf r}') \doteq \exp(-jk|{\bf r}-{\bf r}'|)/4\pi|{\bf r}-{\bf r}'|$ and $k$ is the wave number of the incident field. Imposing the condition that the total pressure $p^t({\bf r}) \doteq p^i({\bf r}) + p^s({\bf r}) = 0$ on the surface $\Omega$ provides an integral equation for the total potential, $\phi^t({\bf r})$, given by  
\begin{equation}
  \phi^i({\bf r}) = \frac{1}{2}\phi^t({\bf r}) - \int_\Omega d{\bf r}' \phi^t({\bf r}') \hat{\bf n}'({\bf r}')\cdot \nabla' g({\bf r},{\bf r}'). 
  \label{eq:ie1}
\end{equation}  

Further, by imposing that the normal component of the velocity goes to zero on the surface of the scatterer, i.e. $\hat{\bf n}({\bf r}) \cdot {\bf v}^t = 0$, $\forall~{\bf r} \in \Omega$, we obtain the normal derivative of the above integral equation. 
\begin{equation}
  \hat{\bf n}({\bf r}) \cdot \nabla \phi^i({\bf r}) = \int_\Omega d{\bf r}' \phi^t({\bf r}') \hat{\bf n}({\bf r}) \cdot \nabla \hat{\bf n}'({\bf r}')\cdot \nabla' g({\bf r},{\bf r}'). 
  \label{eq:ie2}
\end{equation} We define two integral operators ${\mathcal K}$ and ${\mathcal T}$ as 
\begin{subequations}
  \begin{equation} 
    {\mathcal K} \circ [\phi({\bf r})] \doteq  \frac{1}{2}\phi({\bf r}) - \int_\Omega d{\bf r}' \phi({\bf r}') \hat{\bf n}'({\bf r}')\cdot \nabla' g({\bf r},{\bf r}')
    \label{eq:Kop}
  \end{equation}
  and 
  \begin{equation}
    {\mathcal T} \circ [\phi({\bf r})] \doteq  \int_\Omega d{\bf r}' \phi({\bf r}') \hat{\bf n}({\bf r}) \cdot \nabla \hat{\bf n}'({\bf r}')\cdot \nabla' g({\bf r},{\bf r}')
    \label{eq:Top}
  \end{equation}
\end{subequations}
The two integral equations in \eqref{eq:ie1} and \eqref{eq:ie2} can be combined using a parameter $\alpha$ as follows, in a formulation that guarantees uniqueness in the solution $\phi^t({\bf r})$ \cite{Burton1971}; 
\begin{equation}
  \alpha\phi^i({\bf r}) + (1-\alpha)\hat{\bf n}({\bf r}) \cdot \nabla \phi^i({\bf r}) = \alpha{\mathcal K} \circ [\phi^t({\bf r})] + (1-\alpha){\mathcal T} \circ [\phi^t({\bf r})],
  \label{eq:ie}
\end{equation}
where $\alpha \in (0,1)$. 
Solution of equation \eqref{eq:ie} by the method of moments proceeds by representing the unknown potential $\phi^t({\bf r})$ in a set of spatial basis functions, i.e. $\phi^t({\bf r}) = \sum_n{a_n \phi_n({\bf r})}$, where $a_n$ are unknown coefficients. Substituting this representation into \eqref{eq:ie} and  using Galerkin testing results in a matrix system of the form 
\begin{equation}
  \underbar{\underbar{Z}} \underbar{a} = \underbar{f}, 
  \label{eq:mateq}
\end{equation} where \begin{equation}
  {\underbar{Z}} = \left[Z_{i,j}\right] \doteq  \int_{\Omega \cup \Omega_i} d{\bf r} \phi_i({\bf r}) {\mathcal X} \circ [\phi_j({\bf r})],
  \label{eq:matij}
\end{equation}
and ${\mathcal X} \doteq \alpha{\mathcal K}+ (1-\alpha){\mathcal T}$, $\underbar{a} \doteq [a_i]$ and
\begin{equation}
	\underbar{f} = \left[f_i\right] \doteq \int_\Omega d{\bf r} \phi_i({\bf r}) \phi^i({\bf r}).
\end{equation}
Typical method of moments solutions employ polynomial basis functions defined on a simplicial tessellation of the geometry $\Omega$. These basis spaces rely on mapped polynomial functions defined on each simplex.  In keeping with the goals stated in Section \ref{intro}, our approach to solving this problem is to (i) develop an overlapping local surface parameterization each of which will be the domain of the support of the basis function and (ii) define basis functions on these domains. Insofar as the latter is concerned, it builds upon the framework developed in \cite{Nair2009,Nair2010a,Nair2011,Nair2011a} for piecewise flat domains.

\section{Construction of Locally Smooth Surface Parametrizations \label{sec:SSA}}
Next, we prescribe the construction of domains of support for basis functions; these domains overlap and form an open cover of the scatterer $\Omega$.  For the rest of this section, we will assume that we have a point cloud, a set of normals to the surface at these points and a connectivity map that identifies nearest neighbors for each point. Note, that algorithms such as  ball pivoting \cite{Bernardini1999} may be used to obtain such a connectivity map in linear time. Algorithm \ref{alg:algo1} presents a sequence of tasks in order to construct these patches. The rest of this section will elucidate each of the steps presented therein. 

\begin{algorithm}[H]
  \begin{algorithmic}[1]
  \STATE Subdivide initial primitives into overlapping neighborhoods $\{\Omega_i\}$
  \FOR{each neighborhood $\Omega_i$}
  \STATE Project each neighborhood onto a plane $\Gamma_i$
  \STATE Construct a local coordinate system ($u,v,w$) using the projection plane and its normal
  \STATE Construct GMM patches $\Lambda_i$ as a least squares, polynomial approximation to $\Omega_i$ 
  \STATE Merge or split these patches if necessary 
 \ENDFOR 
\end{algorithmic}
  \caption{(Color online) Outline of patch construction \label{alg:algo1}} 
\end{algorithm}

\subsection{Construction of GMM domains}
\begin{figure}
  \centering
  \includegraphics[width=0.5\textwidth,keepaspectratio=true]{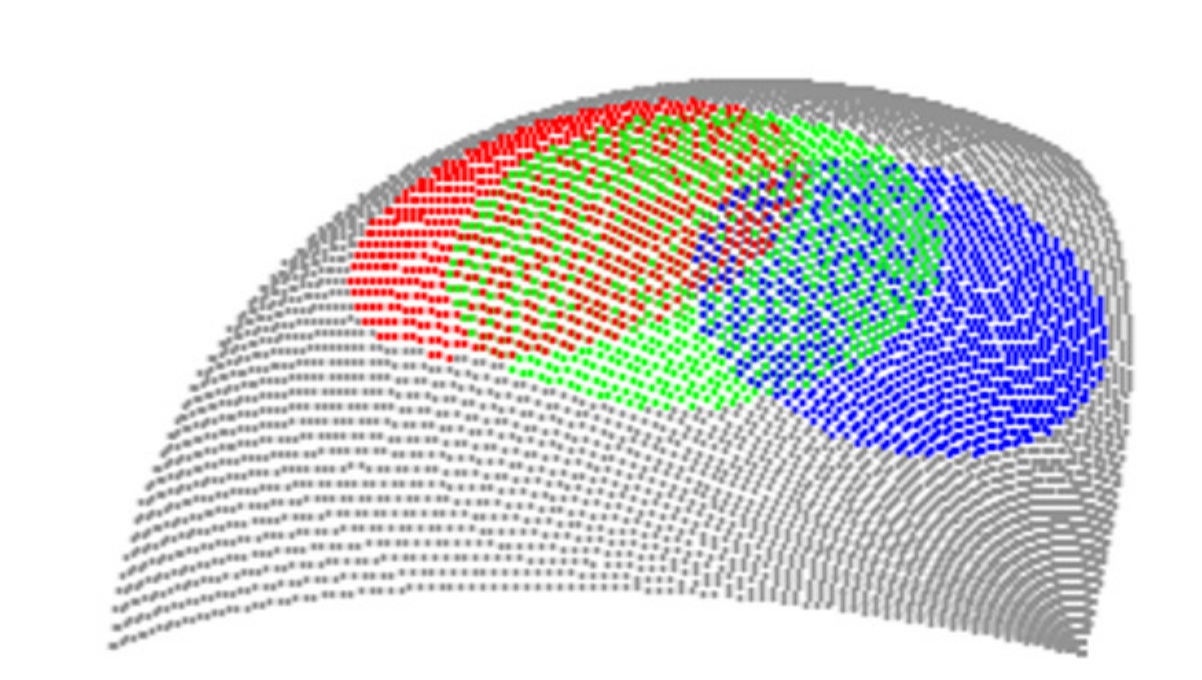}
  \caption{\label{fig:ptpat}(Color online) GMM neighborhoods ($\{\Omega_i\}$) constructed by partitioning a point cloud shown as shaded region. The neighborhoods are constructed as a set of nodes connected to a node.}
\end{figure}

We begin by partitioning the domain $\Omega$ into neighborhoods $\Omega_i$ that overlap and completely cover the domain. To this end, assume that the domain $\Omega$ is described by a set of nodes ${\mathcal N}_L = \cup_{i=1}^{L} \{{\mathcal N}_i\}$,  a connectivity map consisting of primitives $\Delta_N = \cup_{n=1}^N {\left\{ \Delta_n \right\}}$, and finally a unique set of normals $\hat{\bf n}_i$ at these points. Each primitive is defined by a collection of nodes $\Delta_n \doteq \left\{{\mathcal N}_{n,j}\right\}_{j=1}^{j=m_n} \subset {\mathcal N}_L$. In the case of a standard, flat, triangulation, this will reduce to $m_n = 3 \mbox{ } \forall n$, i.e., all the primitives are triangles. To define locally smooth GMM patches, we first start from a collection of primitives that share a node ${\mathcal N}_i$. This collection will be denoted by $\Omega_i$, called the GMM neighborhood. A set of neighborhoods constructed from a point cloud is described in Figure \ref{fig:ptpat}. To construct a locally smooth approximation to $\Omega$ starting from these neighborhoods we first define a neighborhood normal, a projection plane and a notion of permissibility for each neighborhood $\Omega_i$ as follows:
\begin{definition}{Permissible neighborhoods} 

  Given a neighborhood $\Omega_{i}$, centered around a point ${\mathcal N}_i \doteq {\bf r}_i$ and a parameter $\varepsilon$, the average normal for the neighborhood $\Omega_{i}$ is defined as \begin{equation}
    \hat{{\bf n}}_{i,\varepsilon} = \frac{1}{N_i} \sum_{n=1}^{N_i} \frac{1}{m_n} \sum_{k=1}^{m_n} \hat{{\bf n}}({\bf r}_k),
		\label{eq:avnorm}
	      \end{equation} where $N_i$ is the number of primitives $\Delta_n$ connected to ${\mathcal N}_i$,  ${\bf r}_k$ are points chosen on each of the member primitives such that ${\bf r}_k \subset \Omega_{i} \cap \Delta_k$, and $\norm{2}{{\bf r}_k - {\bf r}_i} \leq \varepsilon$. 
	
	      Further, a projection plane for neighborhood $\Omega_{i}$ is defined as the plane passing through ${\bf r}_i$ and normal to $\hat{{\bf n}}_{i,\varepsilon}$. Let $\Gamma_i$ be the projection of $\Omega_i$ on this plane and denote the projection of a point ${\bf r} \in \Omega_i$ to the plane ${\Gamma}_i$ by ${\bf r}'$. The neighborhood $\Omega_{i}$ is    \emph{permissible} if we can find some $\varepsilon$ such that, $\forall {\bf r} \in \Omega_i$ and for ${\bf r} \neq {\bf r}_i$,  \begin{equation}
		\frac{\left({\bf r}-{\bf r}_i\right) \times \hat{{\bf n}}_{i,\varepsilon}}{| \left({\bf r}-{\bf r}_i\right) \times \hat{{\bf n}}_{i,\varepsilon}|} \ge 0.
		\label{eq:local}
	      \end{equation}  
\end{definition}

\begin{figure}
\centering
    \includegraphics[width=0.5\textwidth,keepaspectratio=true]{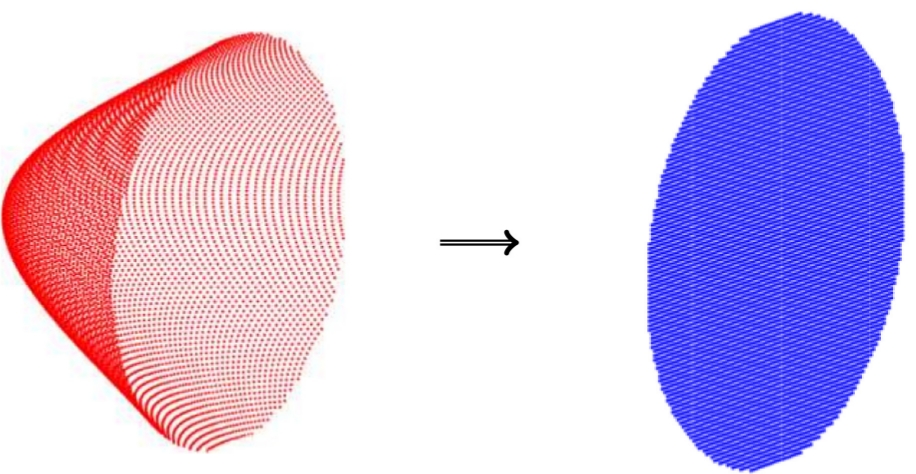}

	\caption{(Color online) Construction of a \label{fig:ptproj}projection plane ($\Gamma_i$) from a GMM neighborhood.}
\end{figure}

\begin{figure}
\centering
\includegraphics[width=0.5\textwidth,keepaspectratio=true]{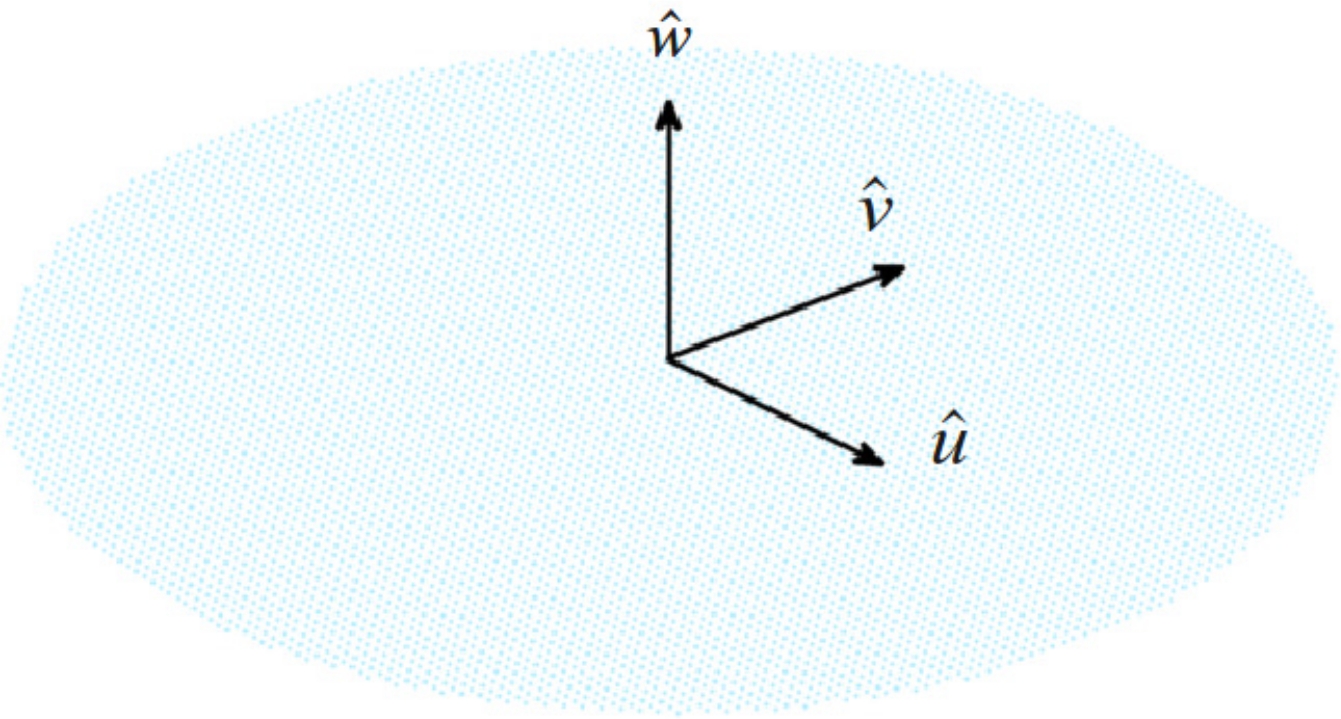}
	\caption{(Color online) Construction of a local coordinate system on the projection plane.}
\end{figure}

In other words, a permissible neighborhood is one for which we can find a projection plane such that the entire neighborhood lies on one side of the plane and has a unique projection on the plane. Figure  \ref{fig:ptproj} shows the construction of the neighborhood normal and projection plane for a permissible neighborhood. For each permissible neighborhood, we can define a local coordinate system containing the projection plane and the neighborhood normal, as follows:

\begin{definition}{Local coordinate system} 
For a permissible neighborhood $\Omega_{i}$,  choose a point ${\bf r}'_m$ such that $|{\bf r}'_m-{\bf r}'_i| > 0$, on ${\Gamma}_i$ and define the following local co-ordinate system $\{\hat{\bf u},\hat{\bf v},\hat{\bf w}\}_i$ and corresponding projections ${\bf u}({\bf r}),{\bf v}({\bf r}),{\bf w}({\bf r})$  for any point ${\bf r} \in \Omega_{i}$ as $\hat{\bf w}  \doteq  \hat{\bf n}_i$, $\hat{\bf u} \doteq {\left({\bf r}'_m-{\bf r}'_i\right)}/{|{\bf r}'_m-{\bf r}'_i|}$, $\hat{\bf v} \doteq {\left(\hat{\bf n} \times \hat{\bf u}\right)}/{|\hat{\bf n} \times \hat{\bf u}|}$, ${\bf u}({\bf r})= {\bf r}' \cdot \hat{\bf u}$, ${\bf v}({\bf r}) = {\bf r}' \cdot \hat{\bf v} $ and ${\bf w}({\bf r}) = {\bf r}' \cdot \hat{\bf w}$. 
\end{definition}

Finally, the above definitions can be used to generate a polynomial map whose domain is the projection $\Gamma_i$, described by the local coordinates $({\bf u},{\bf v})$ and whose range is a smooth surface.  This mapping will become the ``generator'' for the locally smooth surface and is called the GMM surface map.   

\begin{definition}{GMM surface map}
	
  Given a permissible neighborhood $\Omega_{i}$ and a corresponding coordinate system $\{\hat{\bf u},\hat{\bf v},\hat{\bf w}\}$, we can define a polynomial ${\mathcal P}_i^g({\bf u},{\bf v})$ in two variables (${\bf u}$,${\bf v}$) complete to order $g$ by its coefficient vector ${\mathcal C}_i^g \doteq \left[c_0, \ldots c_{(g+1)(g+2)/2}\right]$.   The polynomial ${\mathcal P}_i({\bf u},{\bf v})$ (and corresponding ${\mathcal C}_i^g$),  that minimizes the norm $\displaystyle{\min_{{\bf r} \in \Omega_i}\norm{2}{{\mathcal P}_i^g({\bf u}({\bf r}),{\bf v}({\bf r}))-{\bf w}({\bf r})}}$  can be used to define a transformation ${\mathcal L}_i^g$ from $\Omega_{i}$ to $\Lambda_i$, given by \begin{equation}
		{\mathcal L}_i^g ({\bf r}) : \Omega_i \rightarrow \Lambda_i \doteq {\bf u} \hat{\bf u} + {\bf v} \hat{\bf v} + {\mathcal P}_i^g({\bf u}, {\bf v})\hat{\bf w}.
		\label{eq:surfmap}
	\end{equation} 
	$\Lambda_i$ forms an order-$p$ smooth,  least-squares approximation to $\Omega_i$. This transformation will be called the GMM surface map. The patch $\Lambda_i$ will be called a GMM patch of order $g$. In typical implementations, the ``user'' either chooses a desired order of the patch $g_r$ or an error criteria $\varepsilon_{r}$. Correspondingly the error $\varepsilon_g$ or order $g_{\varepsilon}$ is determined by the minimization procedure. The error is computed at a random selection of points on the patch.  
\end{definition}
\begin{figure}
  \centering
	\includegraphics[width=0.5\textwidth,keepaspectratio=true]{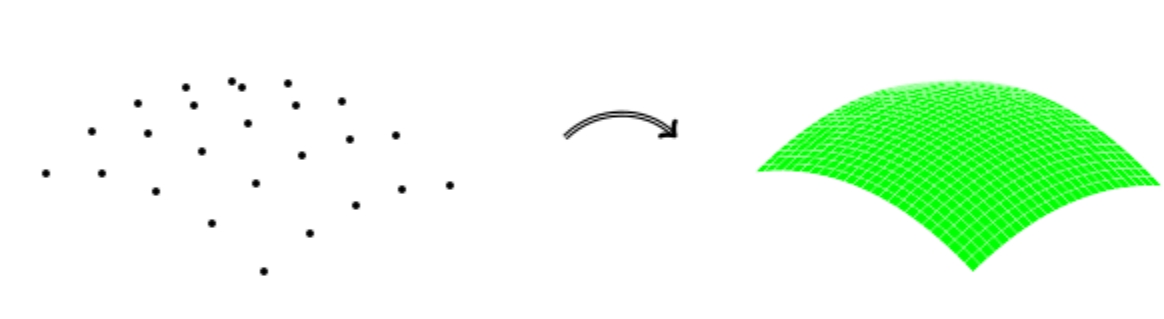}
\caption{\label{fig:smthproj}(Color online) Construction of a locally smooth parameterization starting from a point cloud.}
\end{figure}

\subsection{\label{sec:merg}Merging and splitting of patches}
Next, to achieve complete $h-$ and $g-$adaptivity of the surface parametrization, it is necessary to construct a scheme to merge or subdivide patches as necessary. In this section we will detail ``adaptive'' algorithms to merge or split patches based on a ``sharpness'' criterion. 

\subsubsection{Merging patches}
Let $\Omega_i$ and $\Omega_j$ be two GMM neighborhoods with average normals $\hat{\bf n}_{i,\varepsilon}$ and $\hat{\bf n}_{i,\varepsilon}$ respectively. The neighborhoods are said to fail the smoothness criterion if $ \hat{\bf n}_{i,\varepsilon} \cdot \hat{\bf n}_{j,\varepsilon} \ge \varepsilon_m$, where $\varepsilon_m$ is a user-determined smoothness threshold. The two smooth neighborhoods are merged such that $\Omega_k \doteq \Omega_i \bigcup \Omega_j$, and a new smooth parametrization $\Lambda_k$ is constructed from $\Omega_k$. 

\subsubsection{Splitting patches}
Let $\Omega_i$ be a GMM neighborhood with average normal $\hat{\bf n}_{i,\varepsilon}$. For each point ${\mathcal N}_k \doteq {\bf r}_k$ in the neighborhood, the neighborhood is said to fail the sharpness criterion at ${\mathcal N}_k$, if $\hat{\bf n}({\bf r}_k) \cdot \hat{\bf n}_{i,\varepsilon} \le \varepsilon_s$, where $\varepsilon_s$ is another user-determined sharpness criterion. If a neighborhood fails the sharpness criterion at ${\mathcal N}_k$, the point ${\mathcal N}_k$ is excluded from the neighborhood $\Omega_i$ and a new neighborhood $\Omega_k$, is constructed using primitives that share ${\mathcal N}_k$. Further, two new GMM patches $\Lambda_i$ and $\Lambda_k$ are constructed using (the new ) neighborhoods $\Omega_i$ and $\Omega_k$.  Note that, 
\begin{enumerate}
  \item Since the merging and splitting are operations constructed at the neighborhood level, these operations can be performed either before or after patch construction. This is important if the aim is to obtain $h-$adaptivity. 
  \item The merging and splitting processes can be recursively repeated until all neighborhoods pass both the smoothness and sharpness criteria. 
  \item The smoothness and sharpness criteria need not be global to the problem; in fact, depending on the level of control desired, these criteria can even be on a ``patch-by-patch'' basis.
  \item If the aim is to obtain a true $hp-$refinement scheme, solutions can be constructed and the patches can be merged/split as necessary to refine the solution.
\end{enumerate}

\subsection{Local derivatives, normals and continuity of functions}
  In order to construct functions and surface derivatives on the locally smooth patches, we need to construct surface gradient tensors. Given the GMM surface map ${\mathcal L}^g_i({\bf r})$, we denote its first metric tensor by 
  \begin{eqnarray}
	  \underbar{\underbar{{\bf G}}}_i \doteq  \left[\begin{array}{cc} 
		  g_{11} & g_{12} \\
		  g_{21} & g_{22} \end{array} 
	  \right]\doteq \left[\begin{array}{cc} 
	  \partial_u{\bf r} \cdot \partial_u{\bf r} & \partial_u{\bf r} \cdot \partial_v{\bf r} \\
	  \partial_u{\bf r} \cdot \partial_v{\bf r} & \partial_v{\bf r} \cdot \partial_v{\bf r} \end{array} 
	  \right].
	  \label{eq:mettensor}
  \end{eqnarray} 
  The corresponding surface differential element is denoted by 
  \begin{equation}
	  dS \doteq \sqrt{g_i} du dv,
	  \label{eq:measure}
  \end{equation} 
  where $g_i \doteq det(\underbar{{\bf G}}_i)$, the determinant of the metric tensor. Each term in the tensor can be defined in terms of the polynomial ${\mathcal P}_i^g({\bf u}, {\bf v})$ as \begin{equation}\begin{split}
	  \partial_u {\bf r} = \hat{\bf u} + \partial_u {\mathcal P}^g_i({\bf u},{\bf v}) \hat{\bf w},\\
	  \partial_u {\bf r} = \hat{\bf v} + \partial_v {\mathcal P}^g_i({\bf u},{\bf v}) \hat{\bf w}.	
	  \label{eq:partials}
  \end{split}
  \end{equation}
  
  Given a scalar function $\phi({\bf u},{\bf v})$ defined on the projection plane $\Gamma_i$, the surface gradient of the function on $\Lambda_i$ is given by \begin{equation}
	  \nabla_s \phi \doteq g_{11} \partial_u\phi \mbox{ }\partial_u{\bf r}+  g_{12} \partial_u\phi \mbox{ }\partial_v{\bf r} + g_{21} \partial_v\phi \mbox{ }\partial_u{\bf r} + g_{22} \partial_v\phi \mbox{ }\partial_v{\bf r}.
  \label{eq:surfgrad}
  \end{equation} 
  Higher order derivative tensors on the surface can be described in a similar manner. Finally, the surface normal at any point on the GMM patch can be defined as \begin{equation}
    \hat{\bf n}_i({\bf r}) \doteq \frac{\partial_u{\bf r}\times \partial_v{\bf r}}{\vert \partial_u{\bf r}\times \partial_v{\bf r}\vert}.
    \label{eq:surfnorm}
  \end{equation} Note, that this normal is continuous across the entire patch up to one less than the order of the patch $g$.  

  From the definition of the surface gradient above,  it is clear that any function $\phi({\bf r})$ defined on a GMM patch $\Lambda_i$ of order $g$ that supports $p$ derivatives on $({\bf u},{\bf v})$ with $p \leq g$ will support at least $p$ surface derivatives on the smooth patch $\Lambda_i$. This result implies that defining a function of order $p$ on the smooth GMM patch $\Lambda_i$ is equivalent to defining a corresponding function on the projection plane $\Gamma_i$.   This provides an important tool for defining GMM basis functions as described below. 

  \section{Definition of GMM basis functions\label{sec:GMMBas}} 
The next step is the development of basis functions in each of the above patches. Consistent with the central theme of the GMM framework,  we develop a scheme  that permits different orders of polynomials or different functions to be defined on adjacent patches.  It has been shown, for integral equation based solvers \cite{Nair2011,Nair2011a}, that this can be achieved using a product of two functions; (i) a partition of unity (PU) function that provides continuity of the order of this function across overlapping patches and (ii) a higher order function that determines the quality of approximation within a patch. In what follows, we shall briefly discuss each in turn for completeness. Details of development of basis functions can be obtained from \cite{Nair2011,Nair2011a}, with sufficient modification so as to include the general nature of the local surface description. 

\subsection{Definition of partition of unity functions}

\begin{figure}
\centering
	\includegraphics[width=0.5\textwidth,keepaspectratio=true]{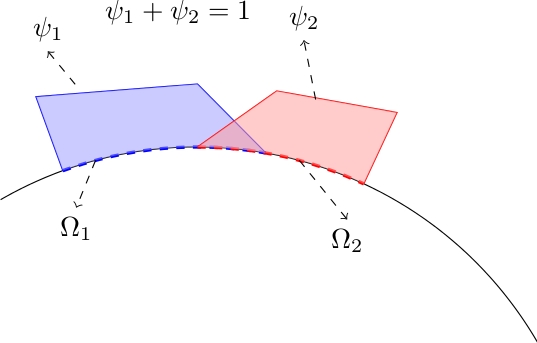}
	\caption{\label{fig:pu1d}(Color online) Definition of a GMM patch and partition of unity}
\end{figure}

\begin{figure}
\centering
	\includegraphics[width=0.5\textwidth,keepaspectratio=true]{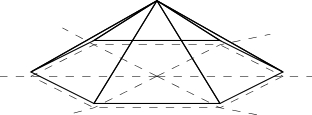}
	\caption{\label{fig:pu2d}(Color online) Definition of a pyramid function for partition of unity -- $\lambda_{i,j}$}
\end{figure}

Consider a GMM patch $\Lambda_0$. Let $\Lambda_i$ overlap with $N_i$ other patches $\displaystyle \{ \Lambda_j\}_{j=1}^{N_i}$. Then a partition of unity function is defined on $\Lambda_i$ as a function $\psi_i({\bf r})$ that satisfies the following properties \begin{enumerate}
  \item $\psi_i({\bf r}) = 0 ~\forall~ {\bf r}   \notin \Lambda_i$
  \item $\psi_i({\bf r}) + \sum_{j=0}^{N_i} \psi_j({\bf r})  = 1 ~\forall~ {\bf r} \in \bigcup \left( \Lambda_i, \left\{ \Lambda_j\right\}_{j=1}^{N_i}\right)$.
\end{enumerate}
In practice, to define a partition of unity function on $\Lambda_i$, we construct a function $\lambda_{i}({\bf u},{\bf v})$ which is $1$ at the patch center and $0$ at the edge of $\Gamma_i$, the projection of $\Lambda_i$. The partition of unity is then defined as 
\begin{equation}
  \psi_i({\bf r})  = \frac{\lambda_{i}({\bf r})}{\sum_k \lambda_{k}({\bf r})}, 
\end{equation} 
where the index $k$ runs through all the patches $\Omega_k$ that overlap with $\Omega_i$. It can be verified that this definition ensures that the partition of unity goes to $0$ at the ends of the patches and adds up to $1$ everywhere on $\Gamma_i$. Correspondingly it satisfies these properties on $\Lambda_i$. Higher order PU functions can be defined in a similar manner if necessary.  For illustration, Figure \ref{fig:pu1d} shows two one-dimensional patches and a partition of unity defined on these patches. Figure \ref{fig:pu2d} shows the construction of $\lambda_{i,j}$ for a flat patch. 


\subsection{Definition of continuous approximation functions\label{sec:proj}}

The next step is to define functions that provide higher order approximation of the unknown field in the patch. As before, we start by defining the function on $\Gamma_i$. Any function $f({\bf u},{\bf v})$ can be now mapped directly to $f({\bf r})$ on $\Lambda_i$.   Note, the domain of the approximation function does not need to be identical to the projection of the patch, $\Gamma_i$. This is possible as functions defined on these patches are eventually multiplied by a PU function that goes to zero at patch boundaries.

One possible choice of approximation functions can be described using Legendre polynomials of the form $\nu_i^m({\bf r}) \in \{P_{p_u}(\tilde{u}){P_{p_v}(\tilde{v})}\}$ where ${P}_q$ denotes a Legendre polynomial of order $q$ and $p_u + p_v \le m$ and
\begin{gather}
  \tilde{u}({\bf r}) \doteq \frac{u({\bf r})}{\max_{{\bf r} \in \Lambda_i} u({\bf r})}, \nonumber \\
  \tilde{v}({\bf r}) \doteq \frac{v({\bf r})}{\max_{{\bf r} \in \Lambda_i} v({\bf r})}.
\end{gather}
Once approximation functions are thus defined, the GMM basis functions are simply products of the approximation function with the partition of unity. That is, 
\begin{equation}
  \phi_i({\bf r}) \in \spn{m}{\psi_i({\bf r})\nu_i^m({\bf r})}
  \label{eq:basfncon}
\end{equation}
Once the basis functions are defined, the next step is the evaluation of the integrals to construct the matrix elements in $\left[Z_{i,j}\right]$. This will be detailed in the next section.  

\section{Evaluation of Matrix Elements\label{sec:mateval}}
The evaluation of the matrix elements in $\left[Z_{i,j}\right]$ involves integrals of the following two forms. 
\begin{eqnarray}
	\int_{\Lambda_i} d{\bf r} \phi_i({\bf r}) \int_{\Lambda_j} d{\bf r}' \phi_j({\bf r}')  \hat{\bf n}'({\bf r}')\cdot \nabla' g({\bf r},{\bf r}')\label{eq:kint}\\ 
	\int_{\Lambda_i} d{\bf r} \phi_i({\bf r}) \int_{\Lambda_j} d{\bf r}' \phi_j({\bf r}') \hat{\bf n}({\bf r}) \cdot \nabla \hat{\bf n}'({\bf r}')\cdot \nabla' g({\bf r},{\bf r}'), \label{eq:tint}.
\end{eqnarray} The integrals need to be evaluated on patches $\Lambda_i$ and $\Lambda_j$. Using the surface differential element defined in \eqref{eq:measure}, we can map the integral of a function $\Theta({\bf r},{\bf r}')$ on a patch $\Lambda_i$  to an integral of the function $\Theta({\bf u},{\bf v},{\bf u}',{\bf v}')$ on the projections $\Gamma_j$ and  $\Gamma_i$ as \begin{equation}
	\int_{\Lambda_i} d {\bf r}	\int_{\Lambda_j} d{\bf r}' \phi({\bf r}) =  \int_{\Gamma_i}\sqrt{g_i} dudv\mbox{ }\int_{\Gamma_j} \sqrt{g_j} du'dv'\mbox{ } \Theta({\bf u},{\bf v},{\bf u}',{\bf v}').
	\label{eq:intmap}
\end{equation}

The evaluation of the integrals in \eqref{eq:kint} and \eqref{eq:tint} are  performed using the transformation in \eqref{eq:intmap} and Gaussian quadrature when the patches are well separated from each other. It is observed the Gaussian quadrature rules converge to sufficient accuracy when the centers of the patches are separated by $d > 0.15\lambda$, where $\lambda$ is the wavelength of the incident field. When the patches are closer to each other, the integrals need to be handled more carefully. We separate the ``near'' evaluations into two cases. \begin{enumerate}
	\item $\Lambda_i$ and $\Lambda_j$ are closer than $0.15\lambda$ but  do not overlap: In this case, the integrals are near singular, but can be evaluated using the techniques described in \cite{Nair2011,Nair2011a,Shanker2010}. 
	\item $\Lambda_i$ and $\Lambda_j$ overlap : In this case, we split the projections $\Gamma_i$ and $\Gamma_j$ into an overlapping section $\Gamma^o$ and two non overlapping sections $\Gamma_i /\Gamma^o$ and  $\Gamma_j /\Gamma^o$. Any integral of the form \eqref{eq:intmap} above can be then re-written as follows \begin{equation} \begin{split}
			\int_{\Gamma_i} dudv \int_{\Gamma_j}du'dv' &= \int_{\Gamma_i /\Gamma^o}dudv\int_{\Gamma_j /\Gamma^o}du'dv' +\int_{\Gamma_i /\Gamma^o}dudv\int_{\Gamma^o}du'dv'\\ &+ \int_{\Gamma_j /\Gamma^o}dudv\int_{\Gamma^o}du'dv' + \int_{\Gamma^o}dudv\int_{\Gamma^o}du'dv'.\end{split} 
			\label{eq:intsplit}
		\end{equation} The preceding equation contains three double integrals that are near singular and one over $\Lambda^o$ that is either singular (for \eqref{eq:kint}) or hypersingular (for \eqref{eq:tint}). The near singular integrals are handled as in case 1 above. To evaluate the singular integrals, we make the assumption that the overlapping portion is locally flat. This implies that $\sqrt{g} = 1$. In this case, it can be shown that the integral in \eqref{eq:kint} reduces to 0. The integral of equation \eqref{eq:tint} on flat patches can be performed by transforming the surface integral into a line integral as described in \cite{Burton1971}. 
\end{enumerate}

\section{Results\label{sec:res}} 

In this section, we present a series of results that demonstrate the features of the GMM scheme presented here. The results can be broadly categorized into two: (i) geometry and function representation, and (ii) using this representation in Galerkin framework to solve integral equations. To start out, we will define error metrics that one can use to define accuracy of reconstruction of a surface. This is then followed by results that demonstrate the following: (i) convergence of surface reconstruction of analytically describable surfaces represented using point clouds; (ii) adaptivity in space and order of surface representation; (iii) convergence of function representation; (iv) the ability of the method to use different basis functions; (iv) ease of $h-$, $p-$, and $hp-$adaptivity, and (vi) the capability of analyzing geometries described using only point clouds (thus obviating the issues with non-conformal tesselations).  In all comparisons, we will use scattering cross section (SCS) data that is obtained with from analytical results or from a over discretized piecewise constant (function and geometry) method of moments solver. 

\subsection{Geometry representation and adaptivity} 

In this section, we deal exclusively with various aspects of representing the geometry using locally smooth functions, as well as $h-$ (space) and $g-$ (order) convergence of these patches. To aid in defining these operations, we define error metrics that will guide adaptivity. 
 
\subsubsection{Error definitions and convergence metrics}

\begin{figure}[!t]
    \centering
    \includegraphics[width=0.45\columnwidth,keepaspectratio=true]{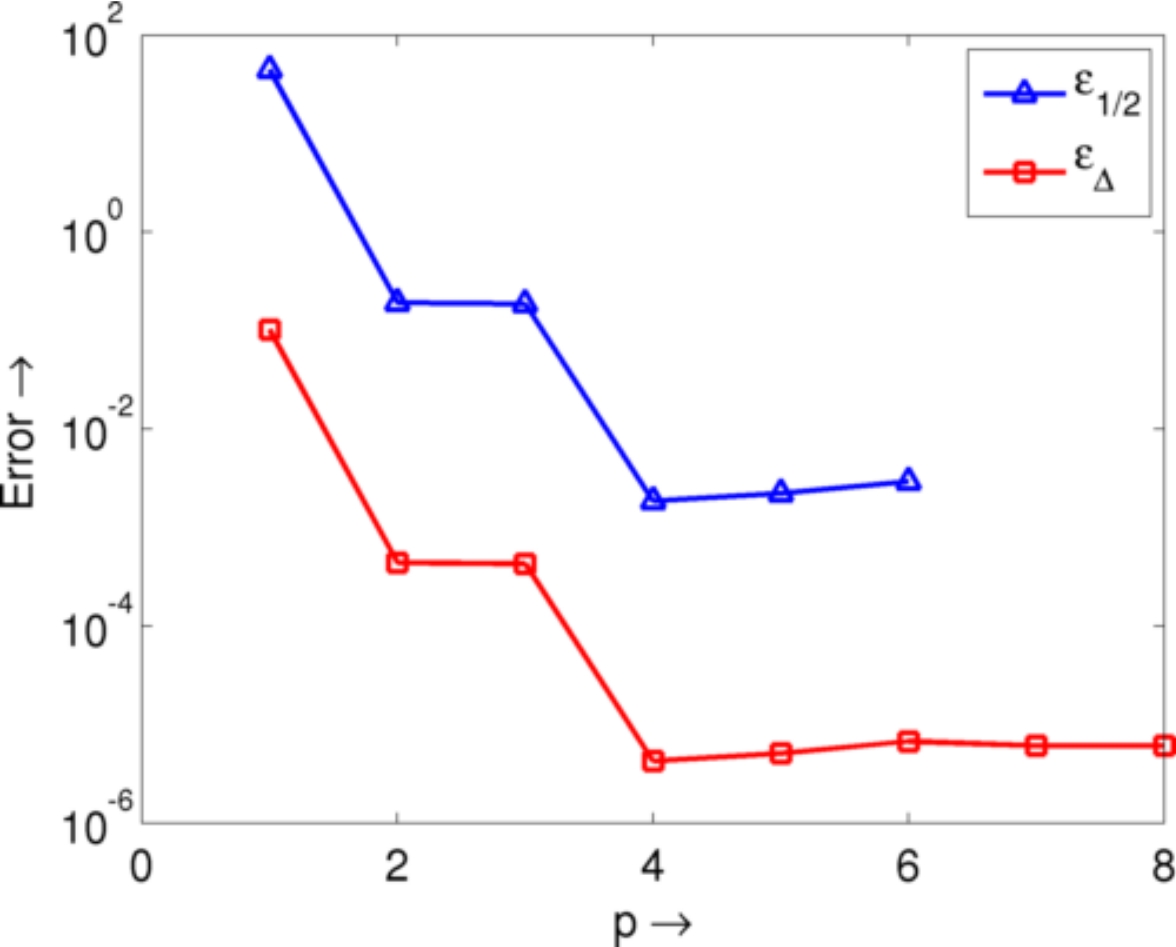}
    \caption{\label{fig:errconv}(Color online) Convergence of surface error metrics with surface patch order.}
  \end{figure}

To begin, we present two error metrics that can be used to determine the quality of representation of a surface. These metrics are suitable for defining functions ${\bf j}({\bf r}) \in H^{1/2}(\Omega)$ on the patch $\Lambda \doteq \bigcup_i \Lambda_i$ and are:
\begin{definition}{Surface Approximation Error}
  Given a surface approximation $\Lambda$ to a true surface $\Omega$, the surface approximation error is defined as 
 {\small \begin{equation}
  \varepsilon_\nabla = \frac{1}{N}\sum_i\left\Vert \Pi_i({\bf r}) \hat{\bf n}_{\Omega}({\bf r}) - \hat{\bf n}_{\Lambda_i}({\bf r})\right\Vert_2; \varepsilon_{1/2} = \frac{1}{N} \left(\sum_i\left\Vert \int_{\Omega} \mbox{d} {\bf r}\mbox{ } \Pi_i({\bf r}) t({\bf r}) - \int_{\Lambda_i} \mbox{d} {\bf r} \mbox{ }t({\bf r})\right\Vert_2 \right) + \varepsilon_\nabla 
    \label{eq:errorMetric}
  \end{equation}}
  where $t({\bf r})$ is any test function and $\hat{\bf n}_{\Omega}({\bf r})$ and $\hat{\bf n}_{\Lambda_i}({\bf r})$ are surface normals to $\Omega$ and $\Lambda$ at ${\bf r} \in \Omega$ and ${\bf r} \in \Lambda_i$ respectively;  $\Pi_i({\bf r})$ is defined by \begin{equation}
    \Pi_i({\bf r}) =
    \begin{cases}
      1 & \forall {\bf r} \in \left.\Omega \right|_{\Omega_{qi}}\\
      0 & \text{ else}
    \end{cases}
    \label{}
  \end{equation}
 \end{definition}
Figure \ref{fig:errconv} demonstrates the convergence of this error on the surface of a sphere of radius $1m$. In order to study convergence, a locally smooth parametrization is constructed starting from a two different point clouds. The first, $p_{tri}=1$, corresponds to a distribution of points such that the average separation distance between nearest neighbors is approximately $0.1m$, and second, $p_{tri}= 2$, corresponds to approximately $0.2m$. The errors $\varepsilon_\nabla$ and $\varepsilon_{1/2}$ are examined as a function of the polynomial order of the patch. As is clear from the image, the error converges very rapidly with the order of the local parametrization. 

\subsubsection{Complex geometry representation}

Next, we present results that demonstrate the construction of a surface representation directly from a point cloud. It will become apparent that the techniques presented can be easily modified to create a smooth surface representation when starting from an underlying mesh. To illustrate such a construction, two candidate structures shown are a gyroid and an icosahedral geometry enclosing a sphere. 

\begin{figure}[!t]
    \centering
    \includegraphics[width=1.0\columnwidth,keepaspectratio=true]{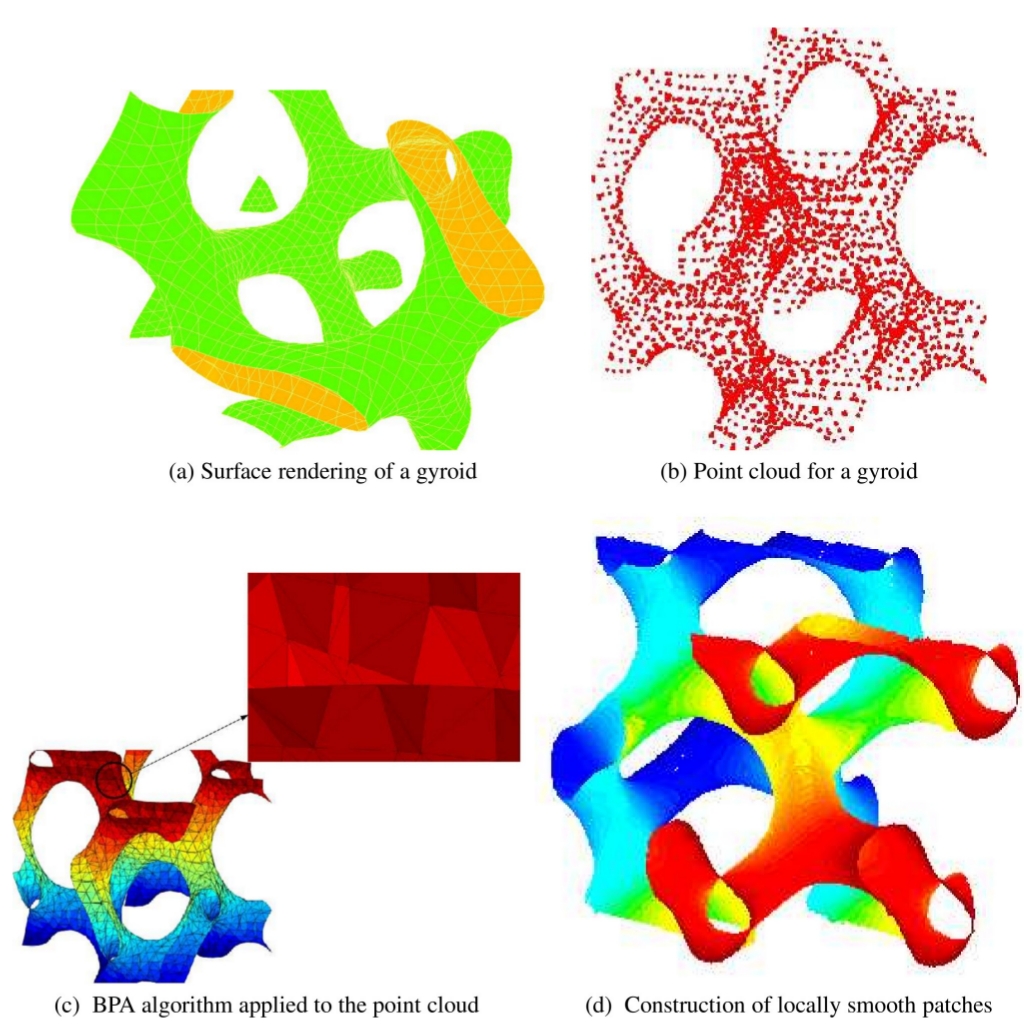}
    \caption{\label{fig:gyr}(Color online) Construction of locally smooth surface representation for a gyroid, starting from a point cloud.}
  \end{figure}

First, Figure \ref{fig:gyr}(a) shows the surface rendering of a gyroid, mathematically described by the equation $\cos(x) \sin(y) + \cos(y)\sin(x) + \cos(z)\sin(x) = 1$. The surface is complex, but as it is analytically known, obtaining a point cloud and corresponding normals at each point is relatively simple. Figure \ref{fig:gyr}(b) shows a point cloud constructed from the gyroid surface description using $-\pi \le \{x,y,z\} \le \pi$ and $N_{pts} = 4888$ points.  Figure \ref{fig:gyr}(c) shows the results of standard meshing algorithm (ball reconstruction \cite{Alliez2011,Rineau2011}) used to create a mesh from the underlying point cloud. As is clear from the inset, the resulting triangulated mesh  has several discrepancies, making it impossible to use in integral equation solvers. Furthermore, even if one were to spend sufficient time in cleaning up mesh, it will result in systems with high condition numbers as the surface discretization is highly non-uniform.  Figure \ref{fig:gyr}(d) shows the surface parametrization algorithm that is  described in this paper applied to the gyroid surface. To construct the parametrization, a primitive is constructed at each point in the point cloud and smooth patches are constructed to an relative error threshold of $\varepsilon_r = 10^{-3}$.  As is clear from the figure, it is possible to obtain a locally smooth parametrization of the surface starting from a simple point cloud. 
  
\begin{figure}[!t]
    \centering
    \includegraphics[width=1.0\columnwidth,keepaspectratio=true]{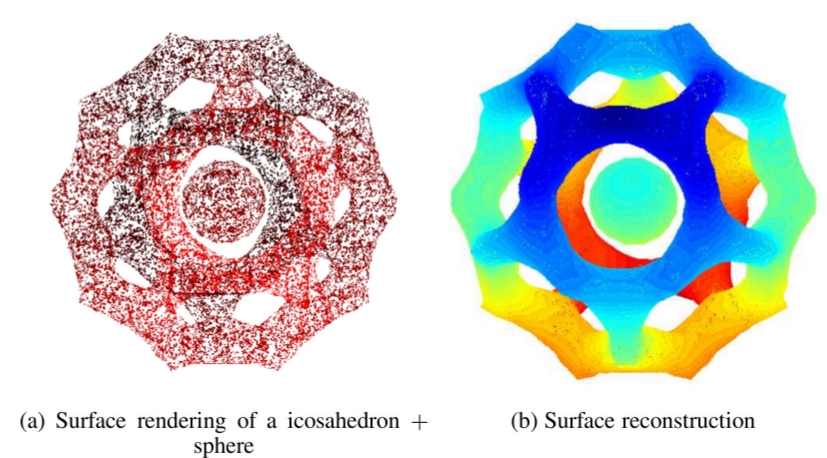}
    \caption{\label{fig:icosa}(Color online) Construction of locally smooth surface representation for an icosahedron $+$ sphere geometry.}
  \end{figure}

  Next, Figure \ref{fig:icosa}(a) shows a point cloud description of an icosahedron enclosing a sphere. The surface of the icosahedron can be constructed in closed form as $cos (x+(1+\frac{\sqrt{5}}{2}y)) + cos(x-(1+\frac{\sqrt{5}}{2}y)) + cos (y+(1+\frac{\sqrt{5}}{2}z)) + cos(x-(y+\frac{\sqrt{5}}{2}z)) + cos (z+(1+\frac{\sqrt{5}}{2}x)) + cos(x-(z+\frac{\sqrt{5}}{2}x))=2$. Figure \ref{fig:icosa}(b) shows a smooth surface parametrization of the surface constructed from the point cloud representation. The point cloud is constructed for an icosahedron of radius $3.0$ and sphere of radius $0.5$ using $N_{pts} = 4328$ points. The final surface is constructed to maintain an relative error of $\varepsilon_r = 10^{-3}$. 

\begin{figure}[!t]
    \centering
    \includegraphics[width=0.45\columnwidth,keepaspectratio=true]{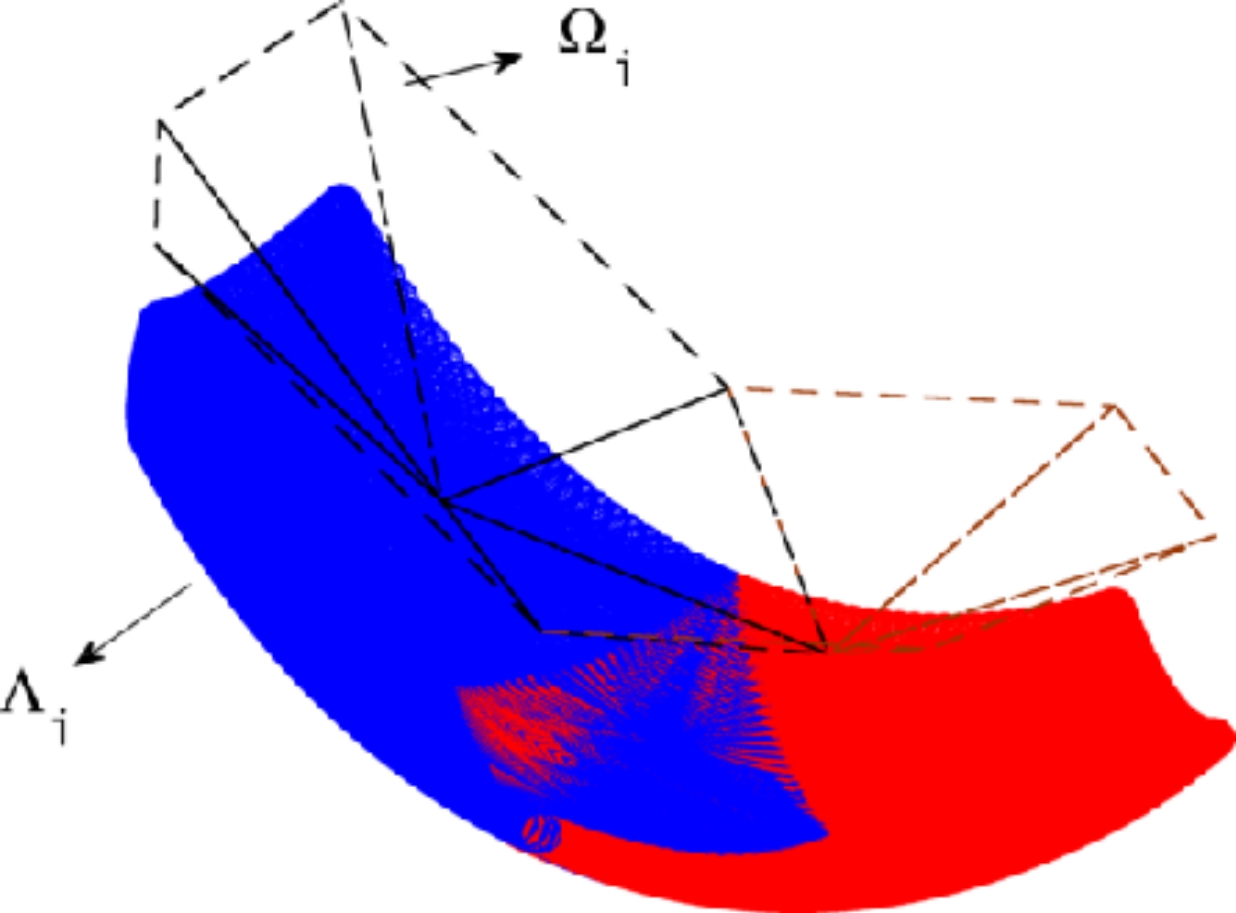}
    \caption{\label{fig:patcon}(Color online) Locally smooth overlapping patches constructed from piecewise smooth neighborhoods.}
  \end{figure}

  Finally, Figure \ref{fig:patcon} demonstrates the reconstruction of two locally smooth GMM patches starting from a piecewise continuous triangulation. Two neighborhoods are defined using $6$ triangles each, shown by $\Omega_i$ and the locally smooth patches $\Lambda_i$ are constructed as approximations to $\Omega_i$. The error in the patches is computed at $300$ arbitrarily chosen points on each triangle making up the patch ($1800$ points overall). In both cases, the relative error is maintained to a threshold of $\varepsilon_r = 10^{-6}$.  Note, that the figure is rendered with an artificial distance between the flat and smooth parametrization for ease of visualization. The three examples provided demonstrate the surface parametrization scheme developed in this paper and its ability to construct patches starting from point cloud descriptions of complex objects and from a standard piecewise triangulation.   

\subsubsection{Automatic adaptivity of geometric order ($g-$adaptivity)}

\begin{figure}[!t]
    \centering
    \includegraphics[width=0.45\columnwidth,keepaspectratio=true]{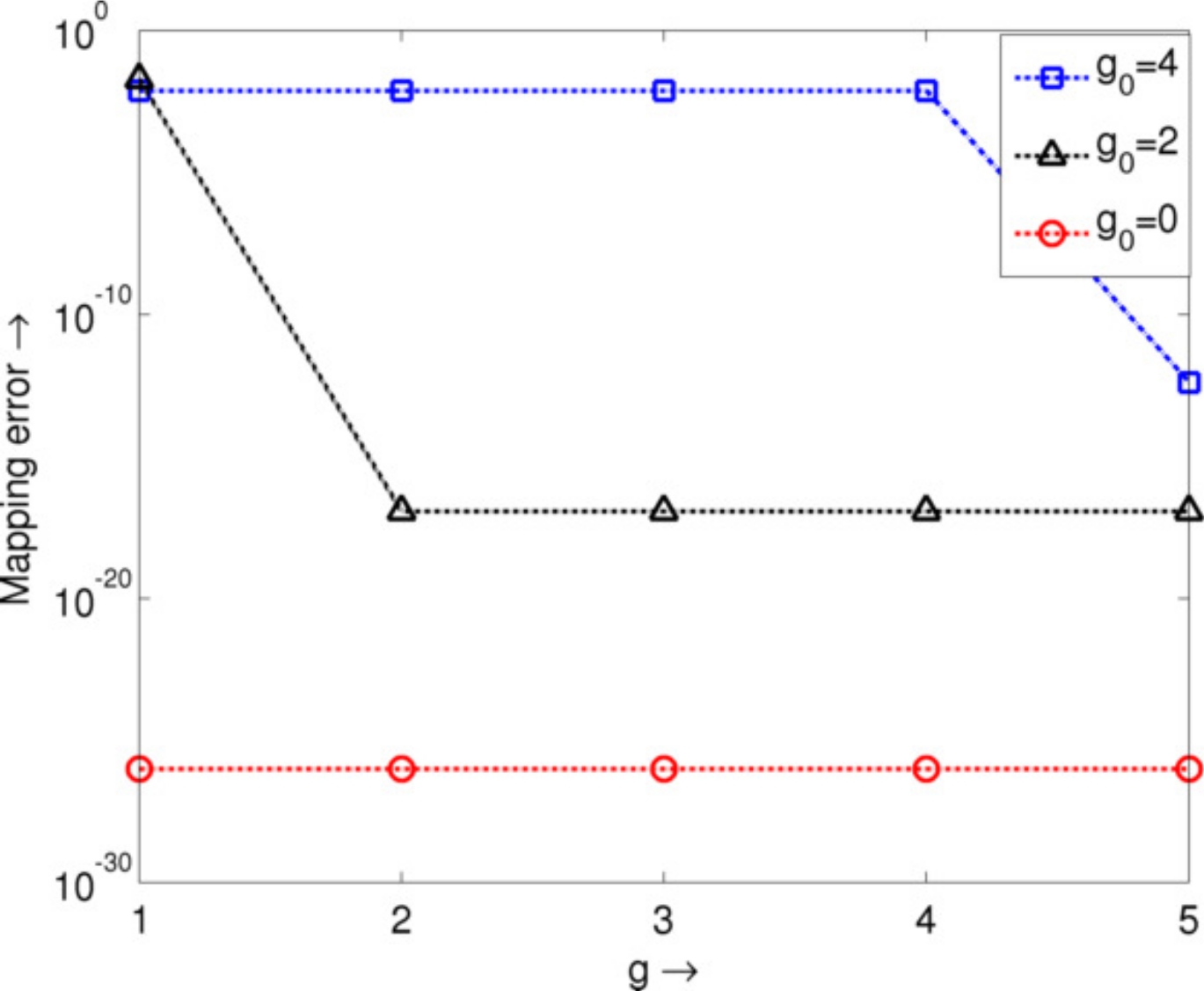}
    \caption{\label{fig:ordconv}(Color online) Convergence of error in normal for surface approximations.}
  \end{figure}

Next,  we demonstrate the polynomial adaptivity of the GMM patch. The use of the least squares minimization, in the construction of the GMM patch, implies that the polynomial order of the map ($g$) can be automatically chosen depending on an error metric, as opposed to being set (by a user) a-priori. To test this property, we consider the error in the surface normal to the reconstructed surface, $\norm{2}{\hat{\bf n}_i({\bf r})-\hat{\bf n}({\bf r})}$. The error is computed with respect to the original surface normal at each point in the neighborhood from which the patch is constructed. Figure \ref{fig:ordconv} shows the error in the surface normal  as a function of the order $g$, for various surfaces, of the form $x^{g_0} + y^{g_0} + z^{g_0} = c$ for a constant $c$ and order parameter $g_0$. In each case, the surface is first approximated using a point cloud description and then GMM neighborhoods $\Omega_i$ are constructed from these triangles. The point clouds are constructed by varying $x$ and $y$ in the interval $-1 \le \{x,y\} \le 1$ and computing $z$ using root finding by an exhaustive search algorithm. In each case, the algorithm is run until $N_{pts} = 625$ valid points are found. 

A locally smooth patch $\Lambda_i$ is then defined for a given order $p$ and the error in the norm is computed. The error convergence in shown for three surfaces, a flat surface (represented as $g_0 = 0$), a piece of a spherical surface ($g_0 = 2$) and a surface with $g_0 = 4$. For the latter two cases $c=16.0$. In each case, the ``true'' surface normal can be computed using the definition of the surface.  As can be seen from the figure, the error for each surface reaches machine precision once the mapping order crosses a threshold. This provides a naturally adaptive mechanism for the choice of surface order. 

  \subsubsection{Merging of patches ($h-$adaptivity)}

\begin{figure}[!t]
    \centering
    \centering
    \includegraphics[width=1.0\columnwidth,keepaspectratio=true]{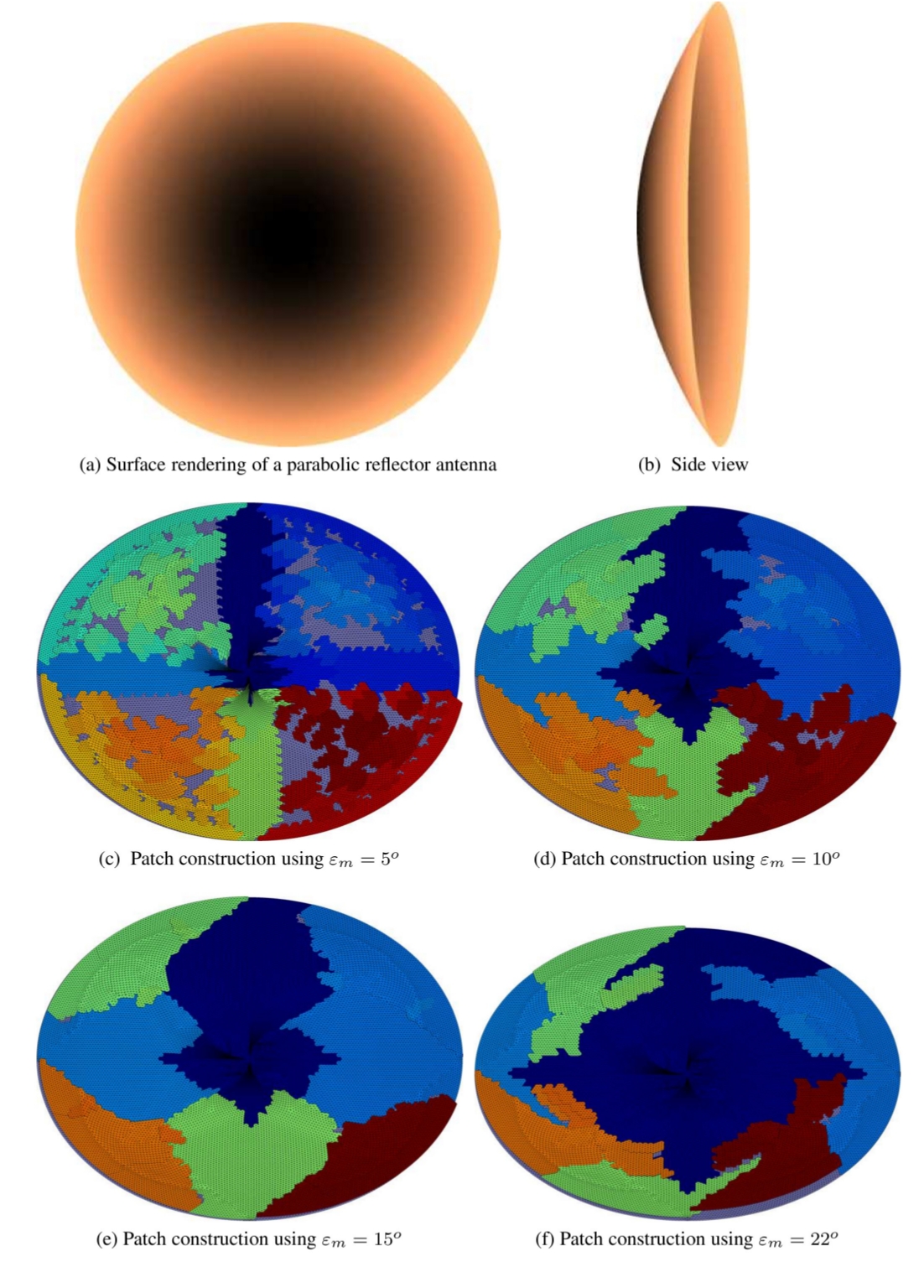}
    \caption{\label{fig:mrg}(Color online) Merging patches for automatic $h-$ adaptivity for a parabolic surface. Patches are shown for different merging criteria.}
  \end{figure}

  Next, we present results on the merging of patches based on a smoothness criterion. Figures \ref{fig:mrg}(a) and \ref{fig:mrg}(b) show two views of a parabolic surface on which a set of triangular neighborhoods is constructed. The surface is obtained by constructing the parabola $y^2 = 4900x$ and then rotating it around the $\hat{\bf x}$ axis. The surface is meshed using $N_0 = 19930$ triangles. The GMM algorithm is then used to (i) construct neighborhoods using the nodes provided by the mesh, (ii) merges these neighborhoods using a constraint on the normals as described in Section \ref{sec:merg} above, and (iii) construct patches from these neighborhoods. Figures \ref{fig:mrg}(c) - \ref{fig:mrg}(f) show the patches constructed starting from the same initial triangulation, using three different thresholds on the angle between the normals ($ \varepsilon_m = \{5^o, 10^o, 15^o, 22^o\}$). In each case, for clarity of representation, only patches that have been constructed from more than $500$ merged primitives are shown. The patches are constructed to maintain a maximum order of $g_r = 2$ .  Table \ref{tab:mrgpatch} shows the final number of patches ($N_{pat}$), the maximum order of patches ($g_{pat}$), the maximum error in the patches $\varepsilon_g$ and number of primitives in the largest patch ($N_{prim}$) for each case in the figure. The figures clearly demonstrate the ability of the GMM scheme to automatically merge neighborhoods and create smooth patches from these neighborhoods.   

  \begin{table}[!h]
    \centering
  \caption{\label{tab:mrgpatch} Patch -statistics for merged patches on parabolic surface}
  \begin{tabular}{|c||c|c|c|c|}
    \hline
    $\varepsilon_m$ & $N_{pats}$ & $g_{pat}$ & $\varepsilon_g$ & $N_{prim}$\\
    \hline
    \hline
    $5^o$ & $827$ & $2$ & $10^{-5}$& $1384$ \\
    \hline
    $10^o$ & $557$ & $2$ & $10^{-5}$& $5315$ \\
    \hline
    $15^o$ & $489$ & $2$ & $10^{-4}$ &$5779$ \\
    \hline
    $22^o$ & $472$ & $2$ & $10^{-3}$& $16614$ \\
    \hline

  \end{tabular}
\end{table}

  \subsection{Function representation}
\begin{figure}
\centering
	\includegraphics[width=0.5\textwidth,keepaspectratio=true]{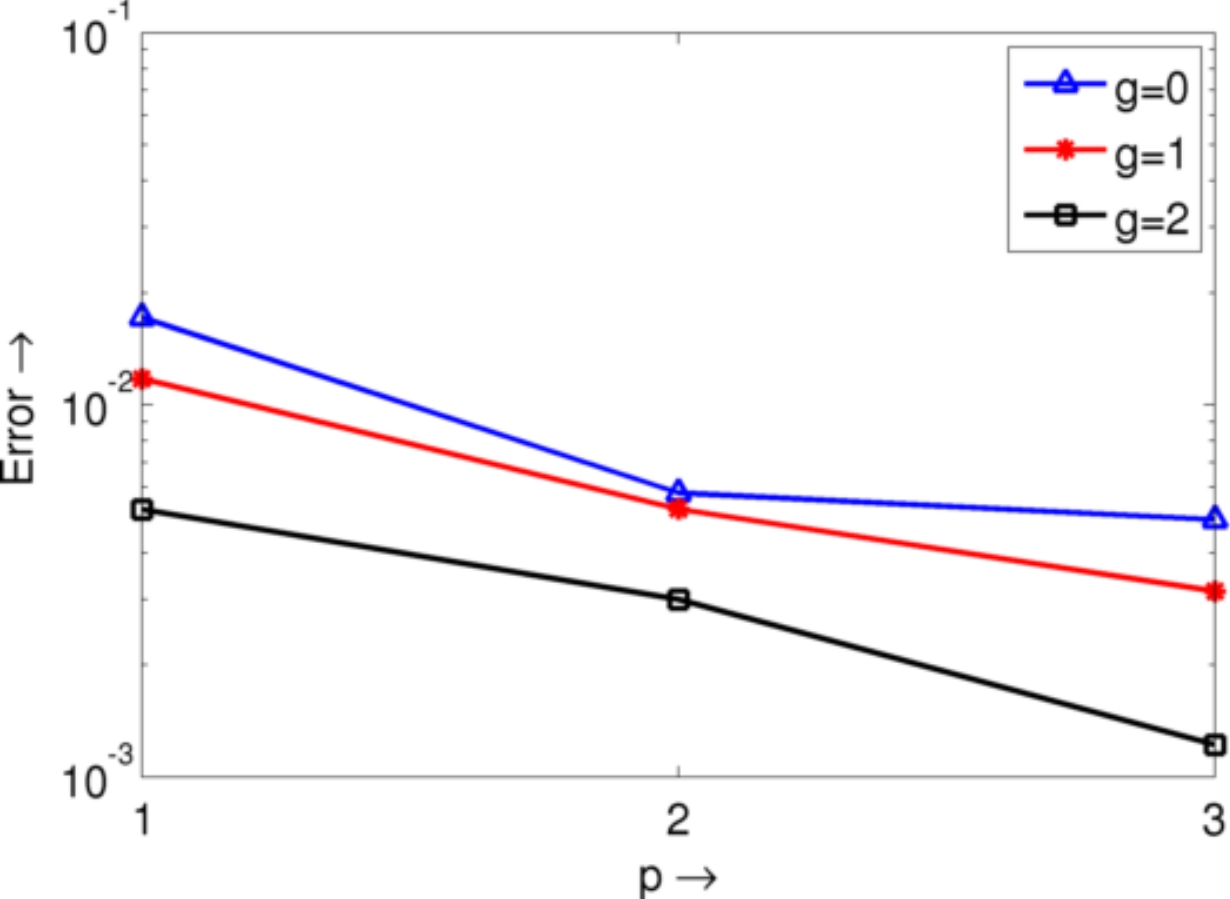}
	\caption{\label{fig:sfn_conv} (Color online) Error convergence for surface functions defined on a sphere.}
\end{figure}

Next, we will demonstrate the ability of the GMM scheme to represent functions on the surface parametrization. Figure \ref{fig:sfn_conv} shows the convergence of the GMM approximation to a  function defined on a spherical surface. To test the efficacy of the GMM basis functions, we define a function of the form $f({\bf r}) \doteq f(\theta,\phi) $. We then construct local surface parametrizations $\{\Lambda_i\}$ of varying order $g = {1,2,3}$ to approximate the surface of the sphere (radius $1.0m$) and construct basis functions of various orders $p={1,2,3}$ on these surfaces. The functions are used to approximate $f({\bf r})$ by setting up the system of equations below.    
\begin{subequations}
  \begin{equation}
  \tilde{f({\bf r})} = \sum_{i}{a_i \phi_i({\bf r})}
  \label{eq:basfn}
\end{equation}
and solving the matrix system resulting from 
\begin{equation}
  \int_\Omega d{\bf r} \phi_j({\bf r}) \sum_i{a_i \phi_i({\bf r})} \approx \int_\Omega d{\bf r} \phi_j({\bf r}) f({\bf r})
  \label{eq:fnapprx}
\end{equation}
\end{subequations}
The coefficients $a_i$ are used to approximate $f({\bf r})$ and the norm of the error on the surface is used as a parameter to test convergence. Figure \ref{fig:sfn_conv} shows the error for $f({\theta},\phi)= \theta + \phi$  as a function of $p$ and $g$, when the patches are constructed from a point cloud on the sphere, consisting of $N_{pts} = 256$ points. A patch is constructed around each point in the point cloud.  As is clear from the figure, the error converges uniformly and logarithmically with both $g$ for each basis function order $p$. 

  \subsection{Application of GMM construct to solving surface IEs}
  
Thus far, we have presented results showcasing the surface parametrization and representation of functions on the surface. Next, we use these basis functions and surface approximations within the Galerkin framework to solve Eqns. \eqref{eq:ie}. To validate the accuracy and utility of the GMM technique implemented on the locally smooth surfaces, we preform a series of numerical experiments. We begin by presenting results that validate the technique on  canonical (or near-canonical) geometries. The data obtained using the GMM is compared against (i) analytical data and (i) a method of moments integral solver that uses flat triangulation and piecewise constant basis functions. Following this, we will present a variety of results that demonstrate (i) $h-$, $p-$ and $hp-$convergence of the GMM scheme, (ii) the ability of the GMM to mix different orders and classes of basis functions and (iii) its ability to handle complex multiply connected geometries, starting from point clouds. Unless specified otherwise, the following is criteria is true for all cases examined: 
\begin{enumerate}
\item The surface representation is constructed using a  point cloud using and a  connectivity map such that the distance between each pair of neighboring points is approximately $0.1\lambda$, where $\lambda$ is the wavelength of the incident field. This allows for comparison against codes constructed on a standard tessellation with average edge length $0.1\lambda$. \item The smooth-surface approximations are constructed starting from the point cloud provided by this discretization. 

\item In each case, the average radius of the smallest circle containing the projection of the GMM patch is used as a measure of the size of the patch, and is maintained to $0.1\lambda$. 

\item Patches are constructed so as to maintain an error threshold of $\varepsilon_r = 10^{-3}$.   

\end{enumerate}
In each case, the bistatic scattering cross section (SCS) is used as a metric for comparison unless otherwise specified. All the cases demonstrated below assume that the test objects are sound-hard and are immersed in a homogeneous medium. The speed of sound in the ambient medium is assumed to be $343 m/s$.

\subsubsection{Validation against analytical results}
First, we consider scattering from acoustically hard spheres of radii $0.1 \lambda$, $0.3\lambda$ and $1.0\lambda$. The incident velocity field has a frequency of $34.3 Hz$, and propagates along $-\hat{\bf z}$ direction. The GMM discretizations, result in $N_{GMM} = {300,450,500}$ unknowns for each of the spheres when using first order Legendre polynomials $p$. In each case, the bistatic SCS evaluated at $\phi=0$ is shown in Figure \ref{fig:sph3}(a) and demonstrates excellent agreement with analytical data.

\subsubsection{$p-$ and $g-$convergence}
\begin{figure}[!t]
    \centering
    \includegraphics[width=1.0\columnwidth,keepaspectratio=true]{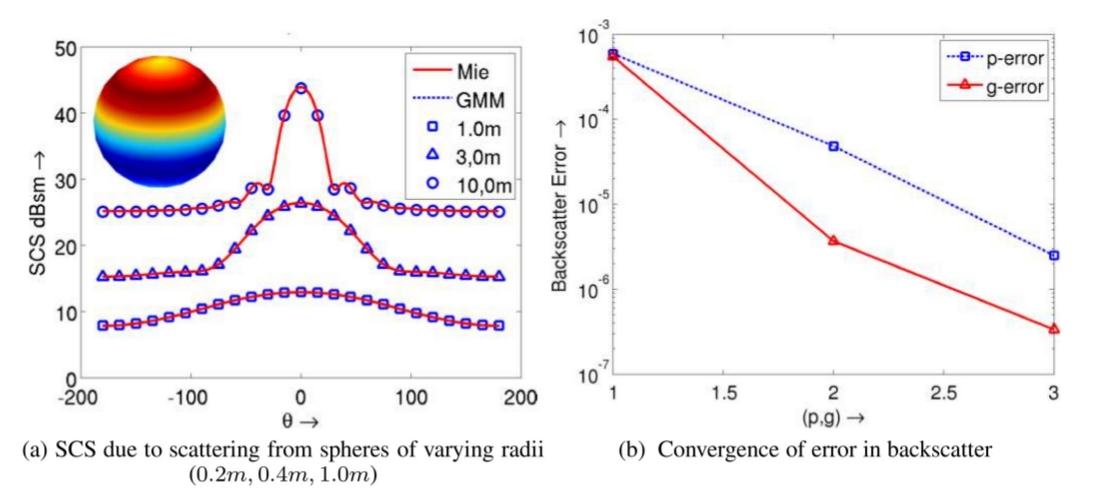}
    \caption{\label{fig:sph3}(Color online) Validation results for the GMM -comparison against analytical results. SCS from a sound-hard sphere is presented as a metric for comparison}
  \end{figure}

Next, we consider  relative error convergence in backscatter from a sphere of radius $0.1\lambda$ between an analytical and the GMM results as a function of  (i) the polynomial order of the basis functions and (ii) the order of local smoothness of the geometry. The incident velocity field is a plane wave of frequency  $343Hz$ propagating along $-\hat{\bf z}$. We consider the convergence of the relative error in backscatter ($\phi=0,\theta=0$). The dashed curve in Figure \ref{fig:sph3}(b) demonstrates $p$ convergence for fixed $h = 0.1 \lambda$ and $g=2$.  The corresponding number of unknowns is $N_{GMM} = {320, 640, 960}$ for $p = 0, 1, 2$, respectively. As is evident from this graph, the error decreases exponentially with increase in $p$. 

Next, the solid line in Figure \ref{fig:sph3}(b) shows the convergence in relative backscatter error with the order of the geometry $g$.  The initial patch size is maintained at $h=0.1\lambda$ and the polynomial order of the basis functions at $p=1$, and as a result the number of unknowns is constant at $N_{GMM} = 320$. As is clear from the figure, the error decreases exponentially with geometry order.

\subsubsection{Validation against piecewise constant MoM}
\begin{figure}[!t]
    \centering
    \includegraphics[width=1.0\columnwidth,keepaspectratio=true]{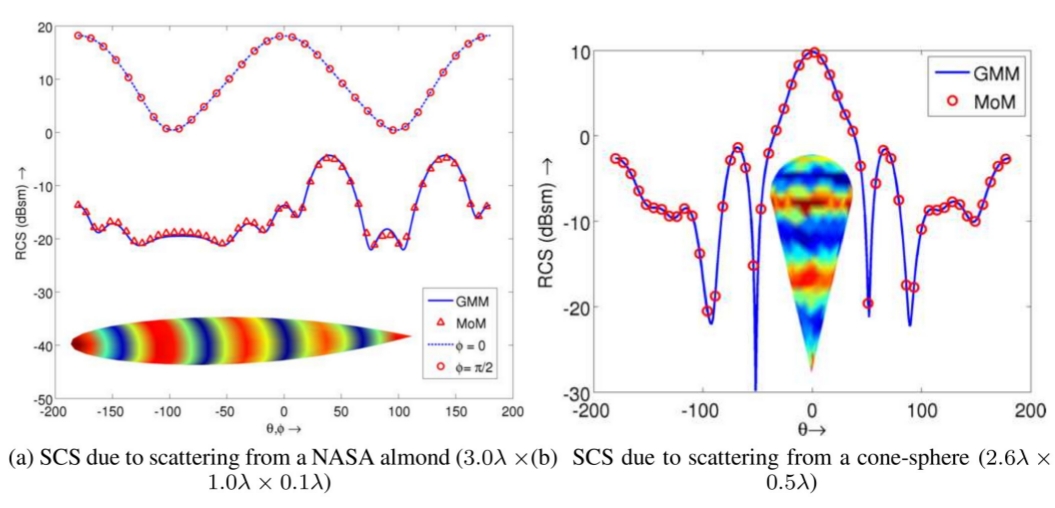}
    \caption{\label{fig:alm-cnsp}(Color online) Validation results for the GMM - comparison against MoM: incident field along $\hat{\bf z}$ for almond, and $\hat{\bf x}$ for conesphere. The SCS evaluated at $\phi=0$ for almond and $\theta = 0$ for cone-sphere. } 
  \end{figure}

Next, we consider two non-canonical geometries - a NASA almond and a conesphere. Figure \ref{fig:alm-cnsp}(a) shows the bistatic SCS (evaluated at $\phi=0$ and $\phi = \pi/2$) due to scattering from a NASA almond, that fits in a box of size $3.0\lambda \times 1.0\lambda \times 0.1\lambda$. A $343Hz$ velocity field is incident along $-\hat{\bf z}$ and the almond is discretized using $N_{GMM} = {1700}$ unknowns. Figure \ref{fig:alm-cnsp}(b) shows the bistatic SCS (computed at $\phi =0$) obtained due to a velocity field incident along $\hat{\bf z}$ on a conesphere with cone-height $2.6\lambda$ and sphere radius $0.5\lambda$. The number of unknowns used to discretize $N_{GMM}={1078}$.  The SCS obtained using the GMM is compared against those obtained using classical MoM.   Excellent agreement is observed between these two data sets. 

\subsection{Mixtures of basis functions}
The results presented thus far validate the GMM scheme and demonstrate  convergence for various parameters ($h$, $p$ and $g$). In what follows, we will present results that demonstrate the ease with which different orders and types of basis functions can be mixed together in the GMM scheme. This capability is thanks in large part to the fact that the basis functions have built in continuity, obviating the need for additional constraints. 
  \begin{figure}[!t]
    \centering
    \includegraphics[width=1.0\columnwidth,keepaspectratio=true]{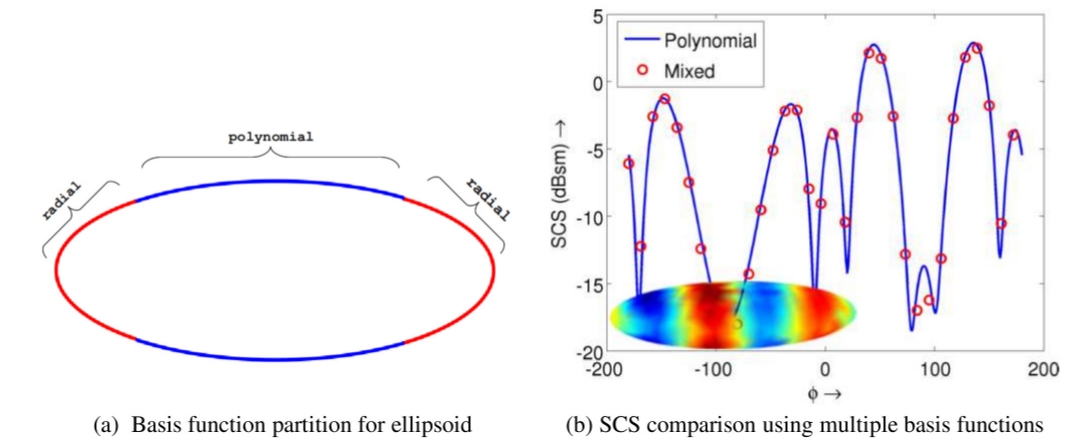}
    \caption{\label{fig:ellipse}(Color online) SCS computed using mixed order basis functions on surface of an ellipsoid.}
  \end{figure}

  First, consider scattering from an ellipsoid of axes $1.0\lambda$, $0.5\lambda$ and $0.25 \lambda$. The ellipsoid is discretized using patches of average radius $0.075\lambda$, and the geometry order is maintained at $g=2$ for all the patches. Polynomial basis functions of order $p=1$ are used in all patches except patches within $0.2\lambda$ of the two ends of the ellipse. In the patches near the end, radial basis functions, inspired by \cite{Golberg1999,Chen2004a} are functions of the form $f(u,v) = \exp{-c_i (u^2 + v^2)}$, where $u,v$ are the local coordinates on the projection plane, as described in \ref{sec:SSA}, and $c_i$ is an order measure, maintained at $c_i =0.5$ for this test.  Figure \ref{fig:ellipse}(a) shows the partitioning of basis functions on the ellipse. Figure \ref{fig:ellipse}(b) shows the SCS obtained using this scheme with mixed basis functions compared against an SCS obtained using polynomial basis functions everywhere ($p=1$), and one using radial basis functions ($c_i = 0.5$) everywhere.  The  SCS is obtained  due to a plane wave incident along $\hat{\bf x}$ and evaluated at $\theta = \pi/2$. Excellent agreement between all these data sets attest to the flexibility of GMM approach. 

\subsection{hp-adaptivity}
\begin{figure}[!t]
    \centering
    \includegraphics[width=0.45\columnwidth,keepaspectratio=true]{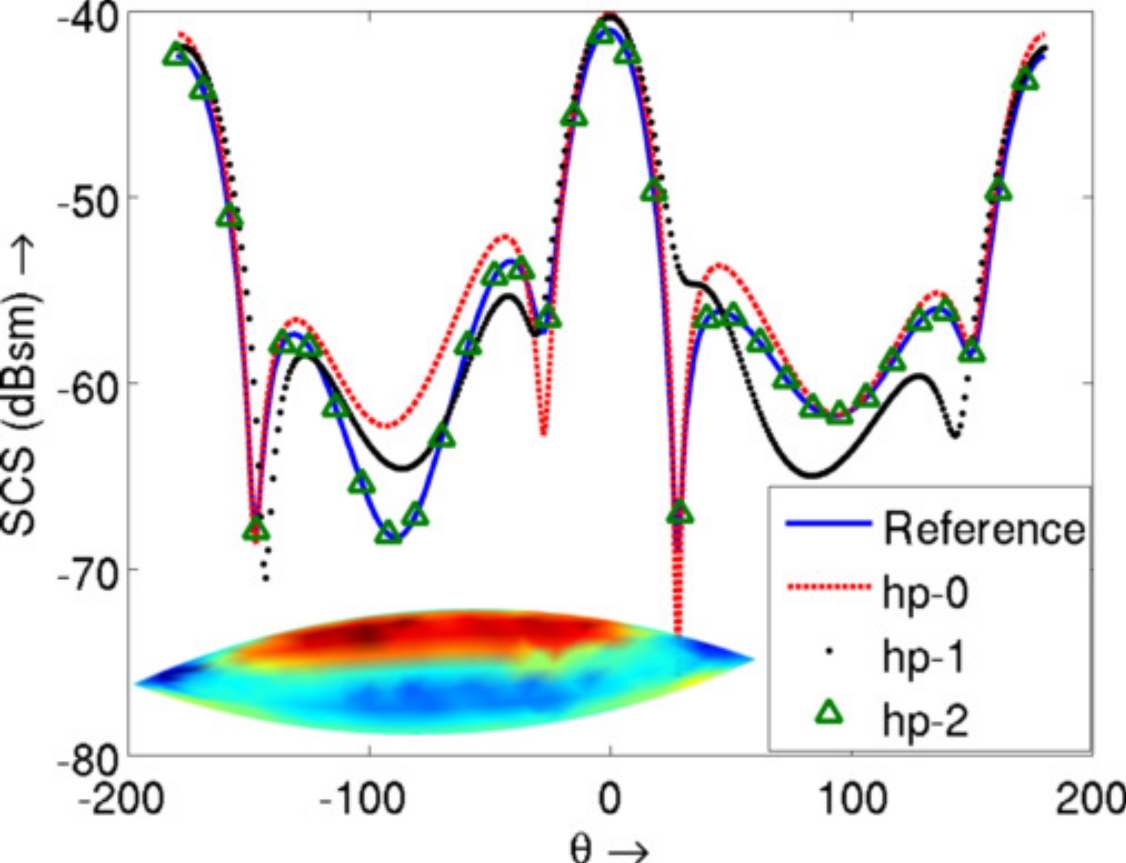}
    \caption{\label{fig:hpconv} (Color online) $hp-$ adaptivity on an ogival surface: Surface currents and SCS comparison for mixtures of various orders of basis functions and patch sizes.}
  \end{figure}

Next, we utilize the flexibility of the GMM scheme to study the $hp-$convergence of the SCS due to scattering from an ogive of size $10m \times 2m \times 10m$. In each of the cases that follows, the SCS is obtained due to a plane wave incident along $\hat{\bf z}$, of frequency $343Hz$. The bistatic SCS is evaluated at $\phi=0$. To obtain a reference, the ogive is discretized using a standard triangulation, at $h=0.05\lambda$ everywhere and the SCS is computed using a classical approach using $N_{MoM} = 8406$ unknowns. The order of surface parameterization used is $g=2$ in the smooth areas, $g=4$ near the ends of the ogive (within $0.25\lambda$ of the end)  and $g=7$ for the two patches near the tips. Using these geometry specifications, the following discretizations are used.  First, the ogive is discretized using patches of size $0.25\lambda$ in the smooth areas and $0.1 \lambda$ near the tips (patches within a sphere of $0.2 \lambda$ near the tips). Basis functions of order $p=1$ are used in all patches resulting in $N_{GMM}=1600$ unknowns. This case is referred to as $hp\! -\!0$ in Figure \ref{fig:hpconv}. Next, basis functions of order $p=1$ are used in the smaller patches and $p=2$ in the larger patches. This case is referred to in Figure \ref{fig:hpconv} as $hp\!-\!1$, and result sin $N_{GMM} = 2156$ unknowns. Finally, the tip of ogive is discretized at $0.1\lambda$, the region near the smooth end of the almond (patches within $0.2\lambda$ of the smooth end)  is discretized at $0.15 \lambda$ and the central, smooth portion is discretized at $0.25\lambda$. Basis functions of polynomial order $p=1$, $p=2$ and $p=3$ are used in each of the areas, respectively. This case is referred to as $hp\!-\!2$, and results in $N_{GMM} = 2876$ unknowns. The agreement of the three different sets of SCS data is shown in Figure \ref{fig:hpconv} and demonstrates the ease with which $hp$ convergence can be obtained using GMM.     
  
  \subsection{Application to objects described using point clouds}
  
\begin{figure}[!t]
    \centering
    \includegraphics[width=0.45\columnwidth,keepaspectratio=true]{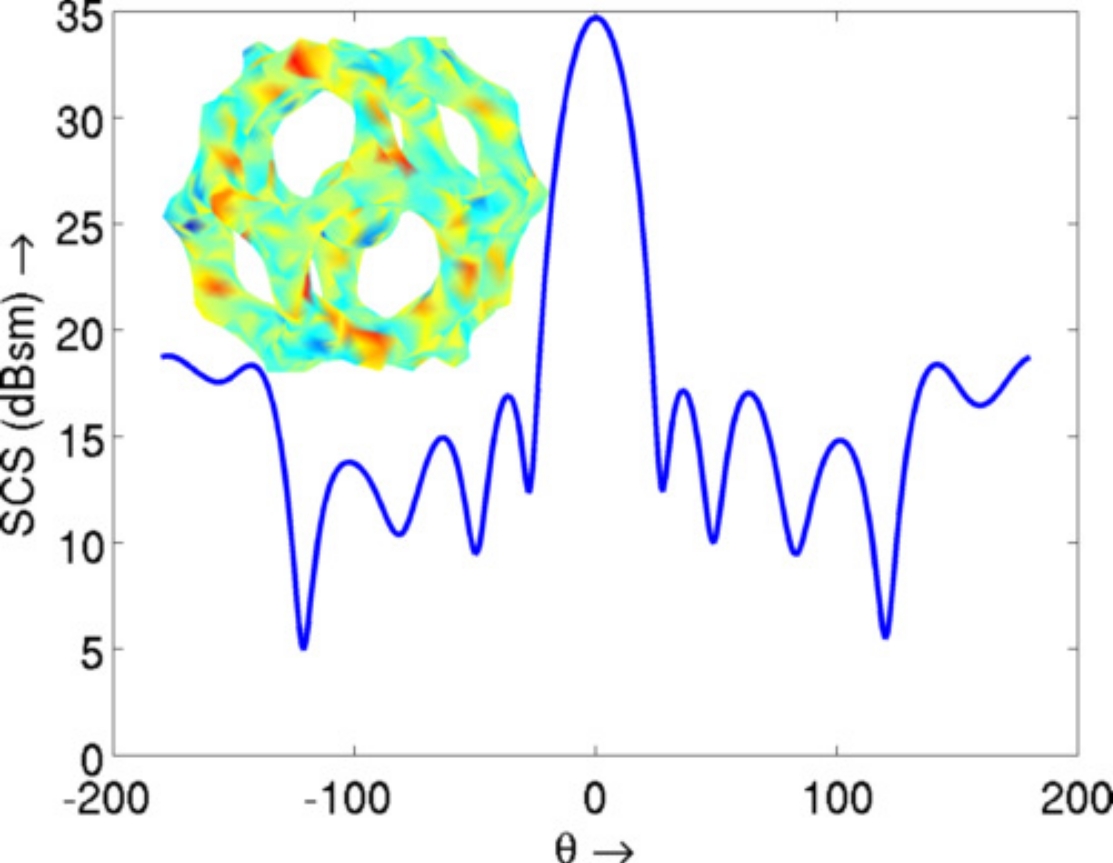}
    \caption{\label{fig:icosares}(Color online)  An application: SCS computation for a complex, multiply connected object (icosahedron enclosing a sphere), initially described using a point cloud.} 
  \end{figure}

  Finally, GMM is used to compute scattering from a complex structure that is difficult to mesh using any standard meshing technique.  Figure \ref{fig:icosares} demonstrates the scattering cross section from an icosahedron enclosing a sphere. This is a complex, disjointed structure. The structure is represented using a point cloud with $N_{pts} = 4888$ points. Patches are constructed from this point cloud with varying $g$, to maintain an relative error of $\varepsilon_r = 10^{-3}$ and basis functions of order $p=2$ are constructed on the patches. The figure shows the surface current on the icosahedron and the SCS computed at $\phi=0$ for an incident acoustic field along $\hat{\bf z}$. 

  \section{Conclusion\label{sec:concl}}
In this paper, we have developed and implemented a highly flexible framework for solving scattering from acoustically hard objects. The framework is very flexible in that the functions used to represent the fields are divorced from the underlying tessellation as continuity is built into representation space. This separation permits relatively easy modification of geometry and function representations, independently, so as to achieve convergence. Here,  several benefits of the GMM has been demonstrated: namely, (i) ability to compute scattering from objects described using either a point cloud or a standard tessellation, (ii) ease of refining the patch size, order of surface, order of approximation and various combinations thereof, and (iii) ability to use mixtures of polynomial and non-polynomial functions. We are currently in the process of developing methods wherein this technique can be applied to solving Maxwell equations, and the results will be presented elsewhere. 

\section{Acknowledgment}

This work was supported by the National Science Foundation (NSF) through grant DMS-0811197 and CCF-1018516. The authors would like to thank the High Performance Computing Center at Michigan State University and the National Center for Supercomputing Applications for the use of computational facilities.
\newpage

\bibliographystyle{jasanum}

\end{document}